\documentclass[10pt,a4paper]{article}
\usepackage[dvips]{color}
\usepackage{epsfig}
\usepackage{amsmath}
\usepackage{graphicx}
\usepackage{authblk}
\usepackage{amssymb,amsmath}
\usepackage{amsmath}

\begin{document}
\title{\bf Interaction between modified Chaplygin gas and ghost dark energy in presence of extra dimensions}
\author{{ M. Khurshudyan$^{a}$ \thanks{Email: khurshudyan@yandex.ru},\hspace{1mm}  J. Sadeghi$^{b}$ \thanks{Email: pouriya@ipm.ir}, \hspace{1mm} M. Hakobyan$^{c, d}$ \thanks{Email: margarit@yerphi.am}\hspace{1mm} H. Farahani$^{b}$ \thanks{Email:
h.farahani@umz.ac.ir}, \hspace{1mm} and R. Myrzakulov$^{e}$ \thanks{Email: rmyrzakulov@gmail.com}}\\
$^{a}${\small {\em Department of Theoretical Physics, Yerevan State
University, 1 Alex Manookian, 0025, Yerevan, Armenia}}\\
$^{b}${\small {\em Department of Physics, Mazandaran University, Babolsar, Iran}}\\
{\small {\em P .O .Box 47416-95447, Babolsar, Iran}}\\
$^{c}${\small {\em A.I. Alikhanyan National Science Laboratory, Alikhanian Brothers St.}}\\
$^{d}${\small {\em Department of Nuclear Physics, Yerevan State University, Yerevan, Armenia}}\\
$^{e}${\small {\em Eurasian International Center for Theoretical Physics,
Eurasian National University, Astana 010008, Kazakhstan}}}  \maketitle
\begin{abstract}
In this paper, we consider three different models of dark energy in higher dimensional space-time and discuss about some cosmological parameters numerically. The first model is a single component universe including viscous varying modified Chaplygin gas. In the second model, we consider two-component universe including viscous varying modified Chaplygin gas and ghost dark energy. In the third model, we consider another two-component universe including viscous modified cosmic Chaplygin gas and ghost dark energy. In the cases of two-component fluids we also consider possibility of interaction between components.
\end{abstract}
\section{\large{Introduction}}
The accelerating
expansion of the universe is the most attractive subject in cosmology. Based on the recent astrophysical data, the universe is spatially flat and
an invisible cosmic fluid, called dark energy with a hugely
negative pressure, may describe this expansion. There are several phenomenological models to describe dark energy. The simplest one is cosmological constant model which has two famous problems called fine-tuning and cosmological coincidence. Also this model don't permit dynamical analysis of universe. In order to suffering these problems, alternative models of dark energy
suggest a dynamical form of dark energy, which at least in an effective level, can originate
from a variable cosmological constant [1, 2], or from various fields, such as a quintessence field [3-8], a phantom field [9-12] or the combination of quintessence and phantom in a unified model named
quintom [13-18]. By using some basic of quantum gravitational principles, one can formulate
further models of dark energy, such as holographic dark
energy [19-25] and agegraphic dark energy models [26].\\
Apart the above models there are interesting
models to describe dark energy based on Chaplygin gas (CG) equation of state. In that case there are several versions to obtain more agreement with observational data, such as generalized Chaplygin gas
(GCG) where it is possible to consider effect of bulk and shear viscosities, as well as consider some constants as variable [27-32]. For example, one can consider varying $G$ and $\Lambda$ in several models of dark energy to obtain a real model of universe in agreement with observational data [33, 34]. The modified Chaplygin gas (MCG) [35, 36] is another version. Recently, viscous MCG is also suggested and studied [37, 38]. A further extension of CG model is called modified cosmic Chaplygin gas (MCCG) with or without viscosity [39-42].\\
Another interesting model of dark energy, which is introduced recently, called Veneziano ghost dark (GD) energy, which supposed to exist to solve the U(1) problem in low-energy effective theory of QCD, and has attracted a lot of interests in recent years
[43-48].\\
On the other hand, higher dimensional space-time introduced in the several physical theories to obtain unified theory [49].\\
In this paper, we consider a two-component fluid [50] with extra dimension as a candidate of universe. We consider the two possibilities for the first component which are kinds of Chaplygin gas. These are viscous varying modified Chaplygin gas and viscous modified cosmic Chaplygin gas. The second component assumed as ghost dark energy. We also investigate possibility of interaction between component. Therefore, we suggest an interacting two-component fluid with dynamic extra dimension as a toy model of the universe.\\
This paper is organized as the following. In next section we introduce our models. In section 3 we write field equations and in section 4 we solve them by using numerical method. In section 5 we summarized our results and give conclusion.
\section{\large{Models}}
In this section we introduce our models and represent properties of them.
In the second and third models we have an interaction term of the following forms,
\begin{equation}\label{eq:NintQ}
Q=(3+d)\gamma H \frac{\rho\rho_{G}}{\rho+\rho_{G}},
\end{equation}
or,
\begin{equation}\label{eq:NintQ}
Q=(3+d)\gamma H\rho_{tot}=(3+d)\gamma H(\rho+\rho_{G}).
\end{equation}

\subsection{\large{First model}}
First of all, we consider the simplest case of single fluid which can be modeled as a viscous varying modified Chaplygin gas given by the following equation of state,
\begin{equation}\label{eq:VCg}
P=A\rho-\frac{B(t)}{\rho^{n}}-(3+d)\xi H,
\end{equation}
where $d$ is the number of extra dimensions and,
\begin{equation}
B(t)=\omega(t)a(t)^{-3(1+\omega(t))(1+n)},
\end{equation}
with $a(t)$ scale factor, $\xi$ is a bulk viscosity coefficient and $H=\dot{a}/a$ is a Hubble expansion parameter. In the ordinary theory the parameter $B$ considered as a constant, but here we assumed it as a variable. For $\omega(t)$ we consider the following parametrization,
\begin{equation}
\omega(t)=\omega_{0}+\omega_{1}t\frac{\dot{H}}{H},
\end{equation}
where $\omega_{0}$ and $\omega_{1}$ are positive constants.
\subsection{\large{Second model}}
In the second model of the work we consider an interaction between a ghost dark energy given by the following energy density,
\begin{equation}\label{eq:GDE}
\rho_{G}=\theta H,
\end{equation}
and a viscous varying modified
Chaplygin gas given by the EoS (1). In the Eq. (4) the parameter $\theta$ is a constant. Therefore, the content of the universe for the second case is an effective fluid with,
\begin{eqnarray}\label{5}
\rho_{tot}&=&\rho+\rho_{G},\nonumber\\
P_{tot}&=&P+P_{G},
\end{eqnarray}
giving EoS parameter as,
\begin{equation}\label{6}
\omega_{tot}=\frac{P+P_{G}}{\rho+\rho_{G}}.
\end{equation}

\subsection{\large{Third model}}
Last model devoted to the interacting ghost dark energy and viscous modified cosmic Chaplygin gas with EoS,
\begin{equation}
P=A\rho-\frac{1}{\rho^{n}}\left [  \frac{B}{1+\omega} -1 + \left ( \rho^{1+n}-\frac{B}{1+\omega}+1\right )^{-\omega} \right ]-(3+d)\xi H,
\end{equation}
where $\omega$ is a constant and a parameter of the model. An advantage of
this model is having stability so the theory is free from unphysical behaviors even when the vacuum
fluid satisfies the phantom energy condition. It is clear that $\omega\rightarrow0$ yields to viscous MCG.

\section{\large{Field equations}}
We use the following FRW metric including extra dimensions,
\begin{equation}
ds^{2}=ds^2_{FRW}+\sum_{i=1}^{d}{b(t)^{2}dx_{i}^{2}},
\end{equation}
where $d$ is the number of extra dimensions, and $ds_{FRW}^{2}$ represents the line element of the ordinary FRW metric in four dimensions which is given by,
\begin{equation}\label{s2}
ds^2_{FRW}=-dt^2+a(t)^2\left(dr^{2}+r^{2}d\Omega^{2}\right),
\end{equation}
where $d\Omega^{2}=d\theta^{2}+\sin^{2}\theta d\phi^{2}$, and $a(t)$
represents the scale factor. The $\theta$ and $\phi$ parameters are
the usual azimuthal and polar angles of spherical coordinates, with
$0\leq\theta\leq\pi$ and $0\leq\phi<2\pi$. The coordinates ($t, r,
\theta, \phi$) are called co-moving coordinates. $a(t)$ and $b(t)$ are the functions of $t$ alone represents the scale factors of 4-dimensional space time and extra dimensions
respectively.\\
The field equations for the above non-vacuum higher dimensional space-time symmetry are [51],
\begin{equation}\label{eq:Fridmman1}
3\frac{\dot{a}^{2}}{a^{2}}=\frac{d}{2}\frac{\ddot{b}}{b} +\frac{d^{2}-2d}{4}\frac{\dot{b}^{2}}{b^{2}}-\frac{d^{2}}{8}\frac{\dot{b}^{2}}{b^{2}}+\rho,
\end{equation}
\begin{equation}\label{eq:Fridmman2}
2\frac{\ddot{a}}{a}+\frac{\dot{a}^{2}}{a^{2}}=\frac{d}{2}\frac{\dot{a}}{a}\frac{\dot{b}}{b}+\frac{d^{2}}{8}\frac{\dot{b}^{}2}{b^{2}}-\frac{d}{8}\frac{\dot{b}^{2}}{b^{2}}-P,
\end{equation}
and,
\begin{equation}\label{eq:fridman3}
\frac{\ddot{b}}{b}+3\frac{\dot{a}}{a}\frac{\dot{b}}{b}=-\frac{d}{2}\frac{\dot{b}^{2}}{b^{2}}+\frac{\dot{b}^{2}}{b^{2}}-\frac{P}{2}.
\end{equation}
Energy conservation $T^{;j}_{ij}=0$ reads as,
\begin{equation}\label{eq:conservation}
\dot{\rho}+3H(\rho+P)=0,
\end{equation}
where Hubble expansion parameter reads as,
\begin{equation}\label{eq:H}
H=\frac{1}{d+3}(3\frac{\dot{a}}{a}+d\frac{\dot{b}}{b}).
\end{equation}
Well known fact is that the interaction between fluid components splits energy conservation equation and for each component we have,
\begin{equation}\label{eq:DE}
\dot{\rho}+(d+3)H(\rho+P)=-Q,
\end{equation}
and,
\begin{equation}\label{eq:DM}
\dot{\rho}_{G}+(d+3)H(\rho_{G}+P_{G})=Q.
\end{equation}
In the next section we use above field equations and investigate cosmological parameters of our models.
\section{\large{Numerical analysis}}
\subsection{\large{Viscous varying modified Chaplygin gas in higher dimensional cosmology}}
Single fluid model can simplify the analyze of the model and sometimes have analytical solution. Taking into account (\ref{eq:conservation}) and (\ref{eq:VCg}), for the dynamics of the fluid we will have,
\begin{equation}\label{eq:sfdynamics}
\dot{\rho}+(3+d)H\left (1+A-\frac{B(t)}{\rho^{n+1}} \right )\rho = (3+d)^{2}\xi H^{2}.
\end{equation}
The solution of the last equation together with the field equations allow us to obtain graphical behavior of all cosmological parameters. Plots of Fig. 1 show that the Hubble expansion parameter of this model is decreasing function of time which yields to a constant at the late time. The first plot of Fig. 1 shows variation with number of dimensions. The green line corresponds to $D=10$ which is interesting in string theory point of view. The second plot shows that variation of $\omega_{0}$ and $\omega_{1}$ is not many important in evolution of $H$. The lost plot shows that increasing viscosity increases value of Hubble expansion parameter.\\

\begin{figure}[h!]
 \begin{center}$
 \begin{array}{cccc}
\includegraphics[width=50 mm]{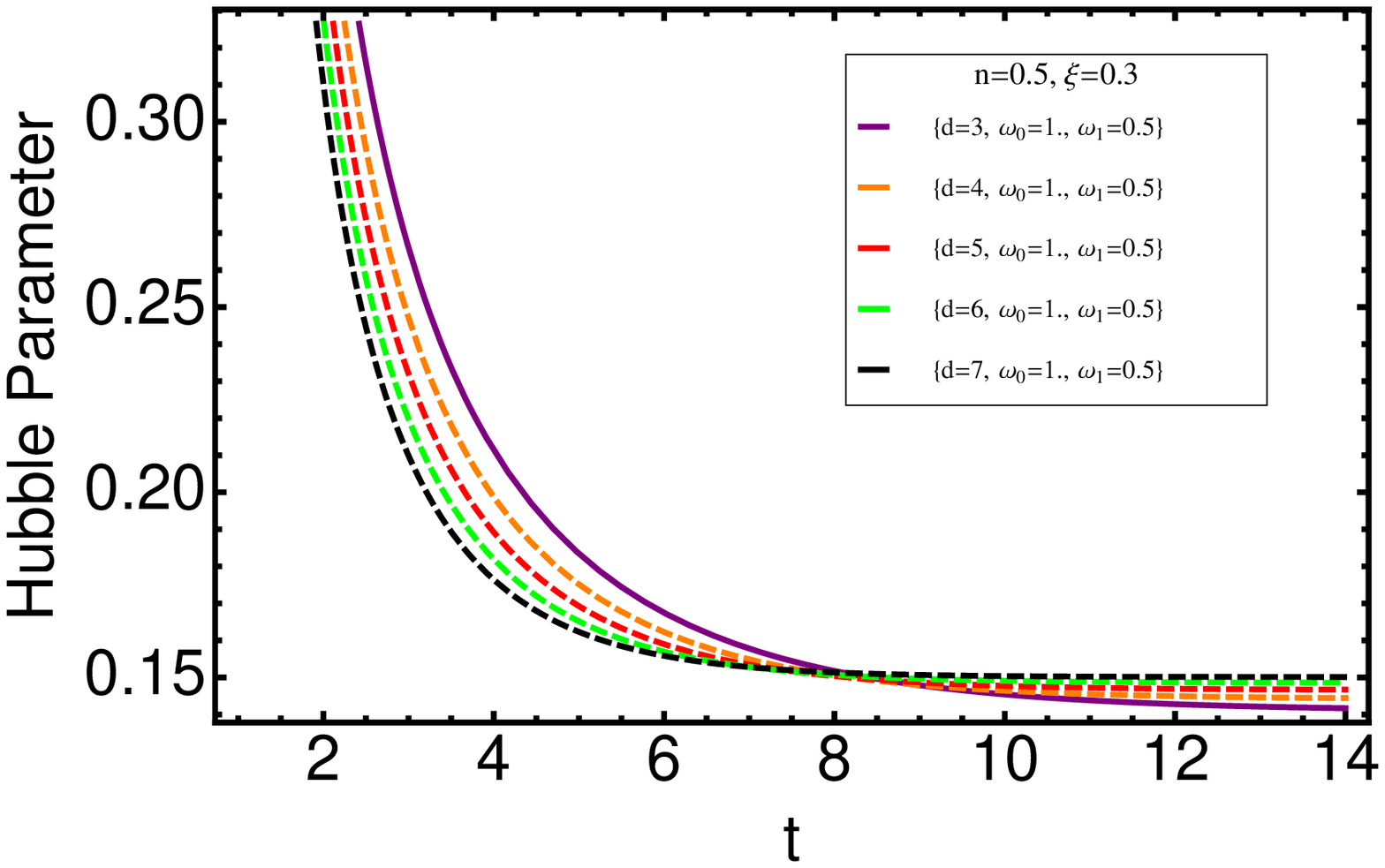} &
\includegraphics[width=50 mm]{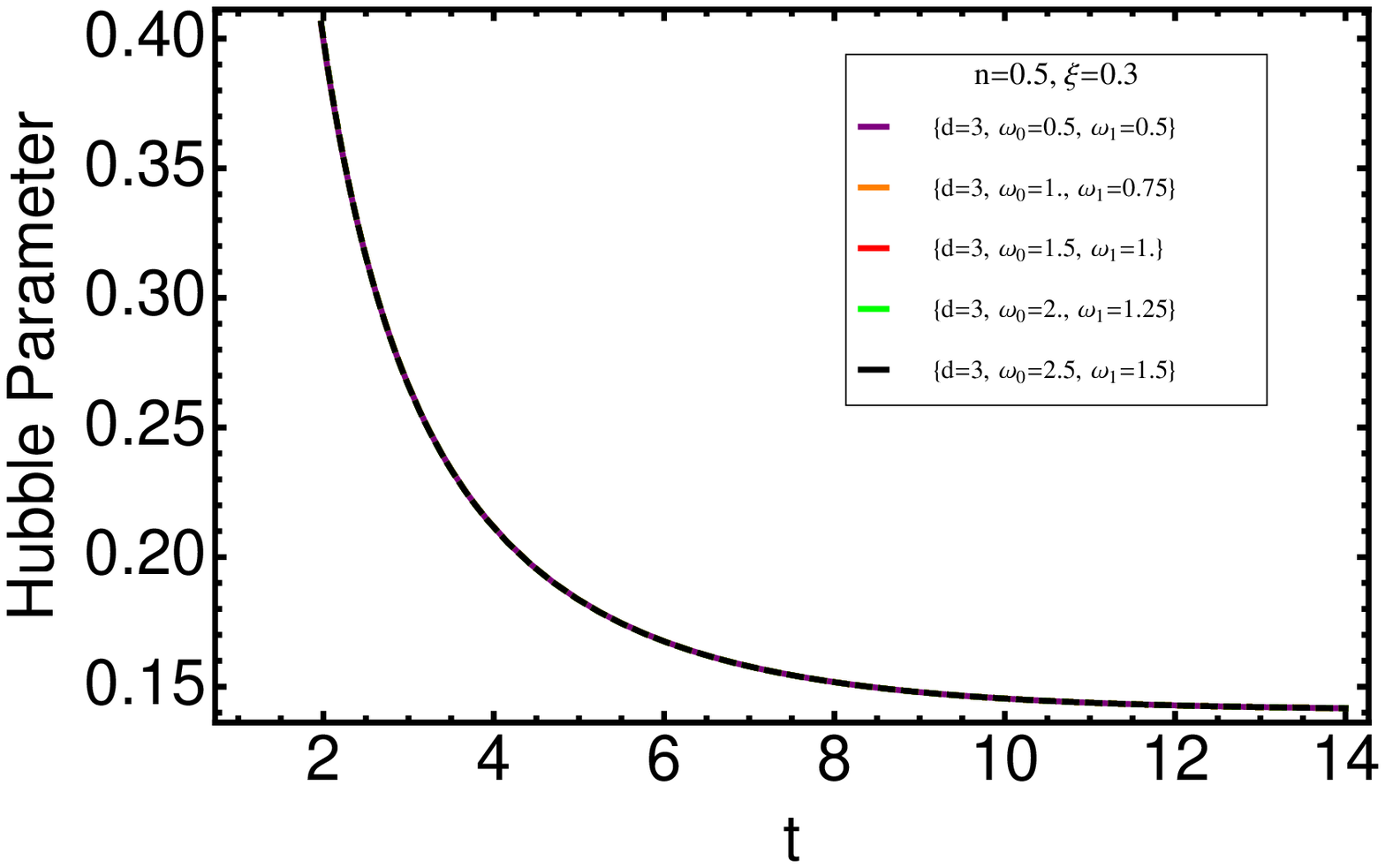}&
\includegraphics[width=50 mm]{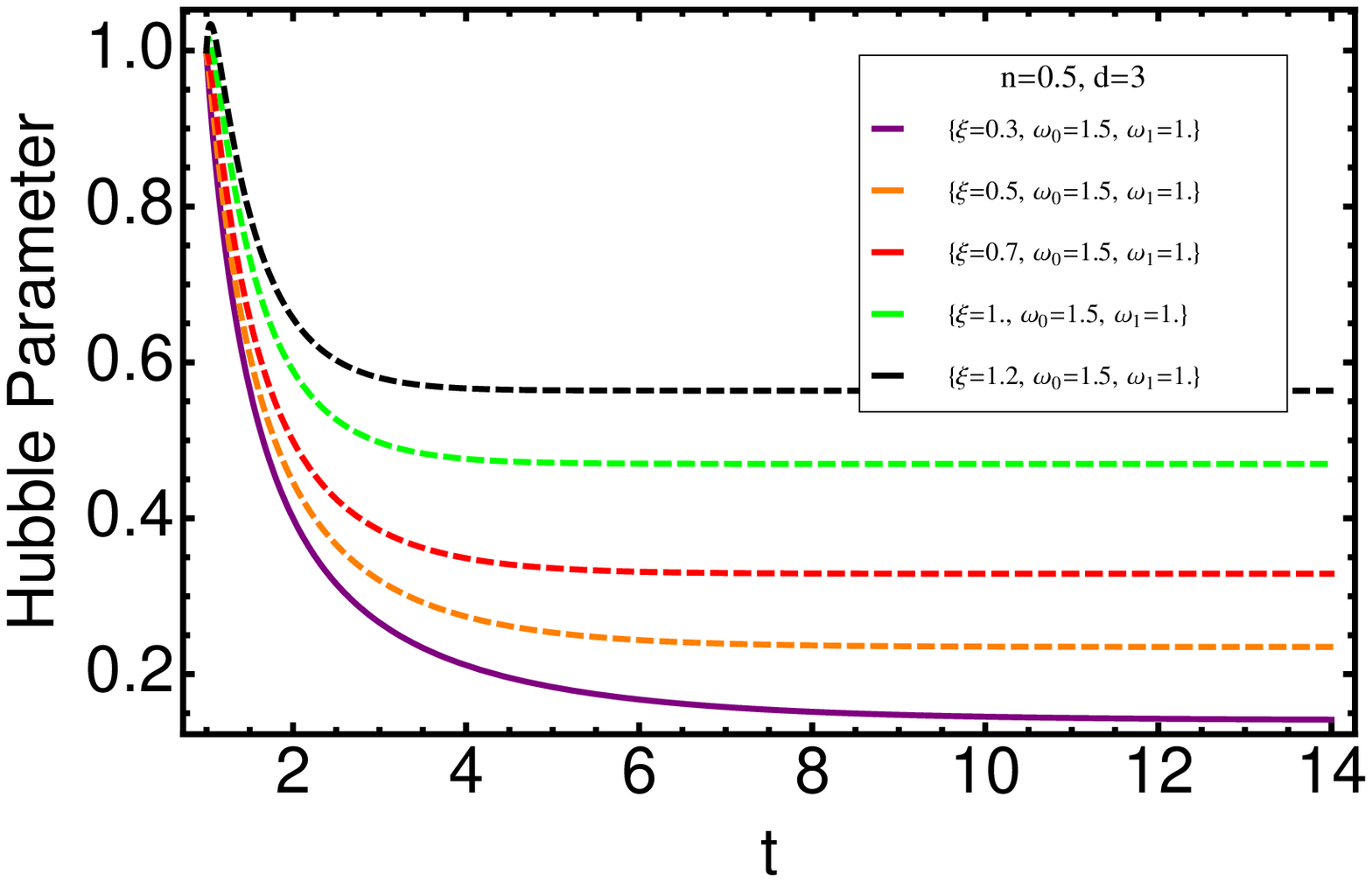}
 \end{array}$
 \end{center}
\caption{Behavior of $H$ against $t$. Single fluid model.}
 \label{fig:1}
\end{figure}

\begin{figure}[h!]
 \begin{center}$
 \begin{array}{cccc}
\includegraphics[width=50 mm]{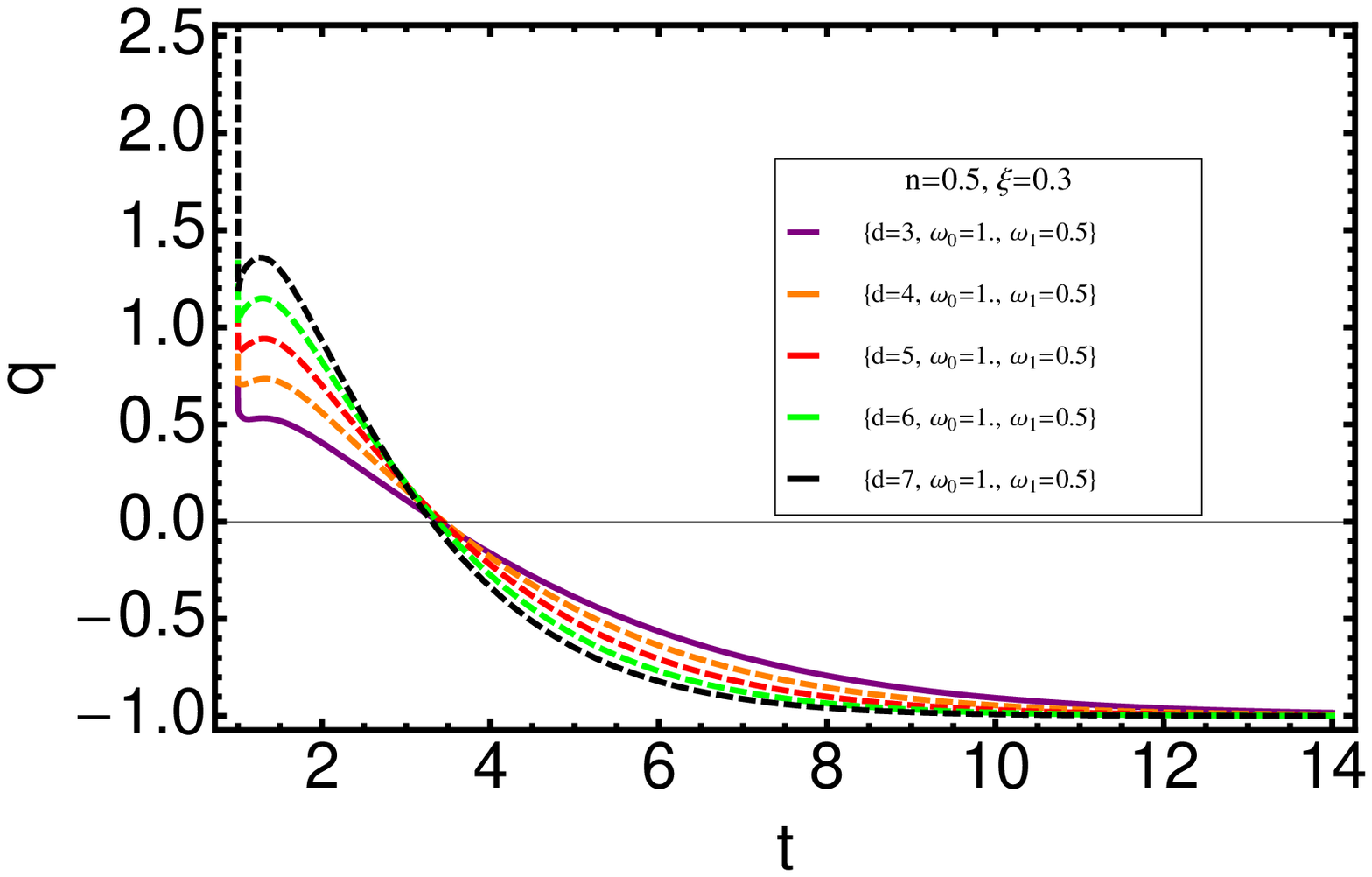} &
\includegraphics[width=50 mm]{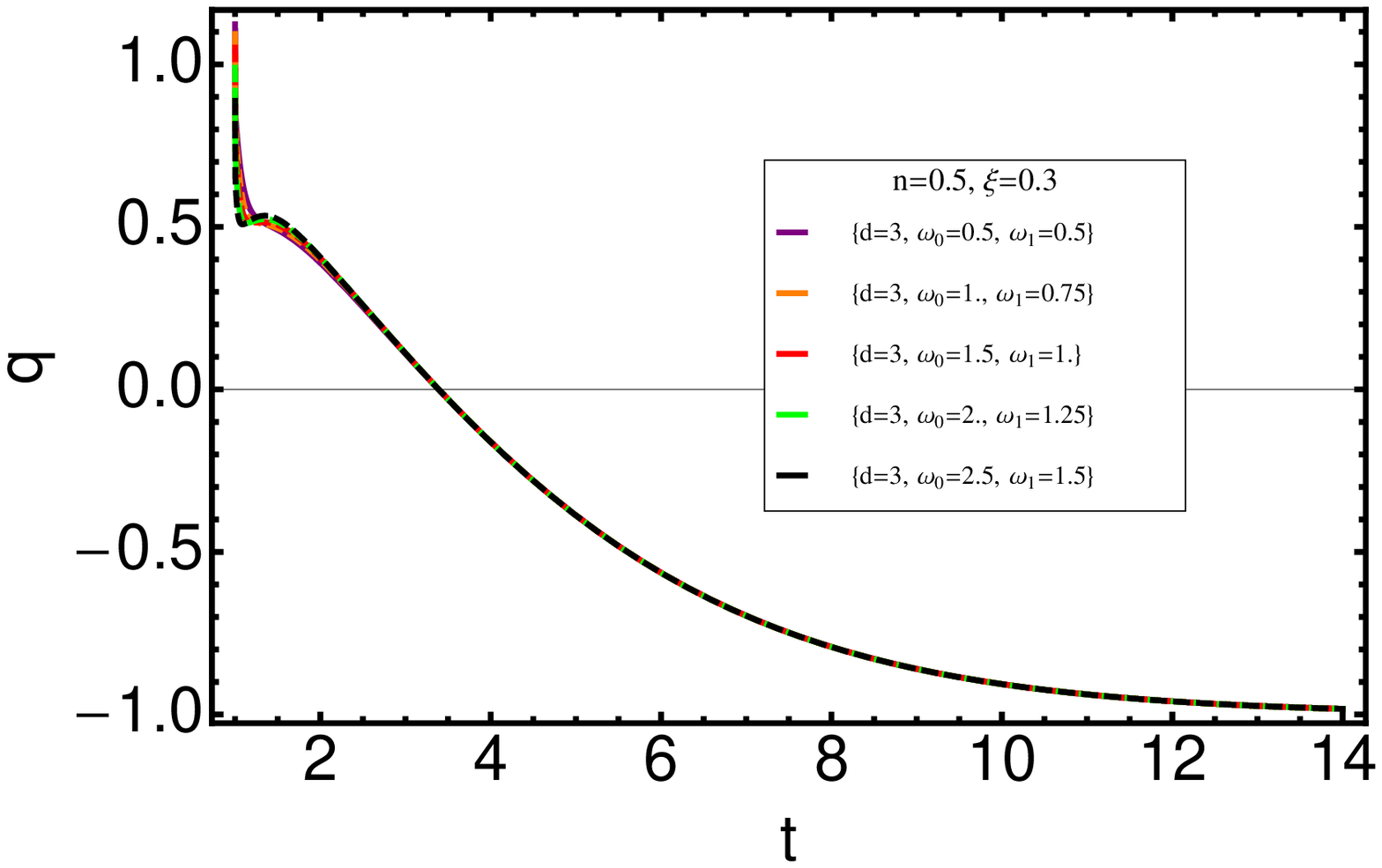}&
\includegraphics[width=50 mm]{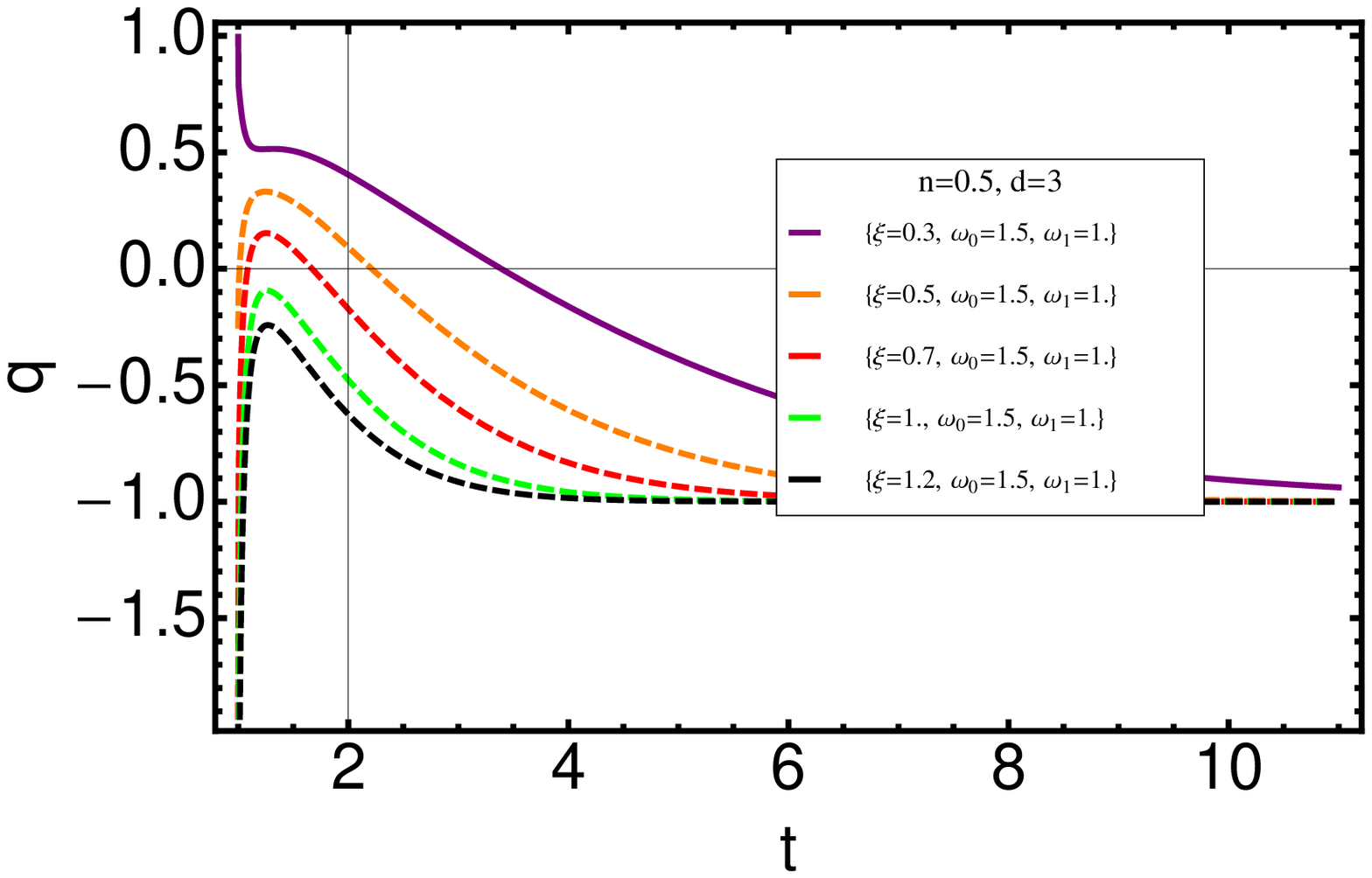}
 \end{array}$
 \end{center}
\caption{Behavior of $q$ against $t$. Single fluid model.}
 \label{fig:2}
\end{figure}

Plots of Fig. 2 show evolution of deceleration parameter,
\begin{equation}\label{20}
q=-1-\frac{\dot{H}}{H^{2}}.
\end{equation}
Interestingly, we can see acceleration to deceleration phase transition for the appropriate values of parameters. The first plot shows that increasing number of dimension increases value of deceleration parameter in the acceleration phase but decreases its value in the late time which is in the deceleration phase. The second plot shows that the variation of $\omega_{0}$ and $\omega_{1}$ is not many important at the late time but makes a little change in the early time. The third plot shows that the value of viscous parameter can't take arbitrary value. In order to have acceleration to deceleration phase transition the value of viscous parameter restricted. For the cases of $\xi\geq0.5$, the universe begun in the deceleration phase which suddenly gone to acceleration phase if $\xi<1$, then translated to the deceleration phase and yields to -1 at the late time in agreement with $\Lambda$ CDM model.\\
In Fig 3 we draw EoS parameter,
\begin{equation}\label{21}
\omega_{Ch}=\frac{P}{\rho},
\end{equation}
and confirmed that $-1\leq\omega_{Ch}\leq-1/3$, which at the late time $\omega_{Ch}\rightarrow-1$.

\begin{figure}[h!]
 \begin{center}$
 \begin{array}{cccc}
\includegraphics[width=50 mm]{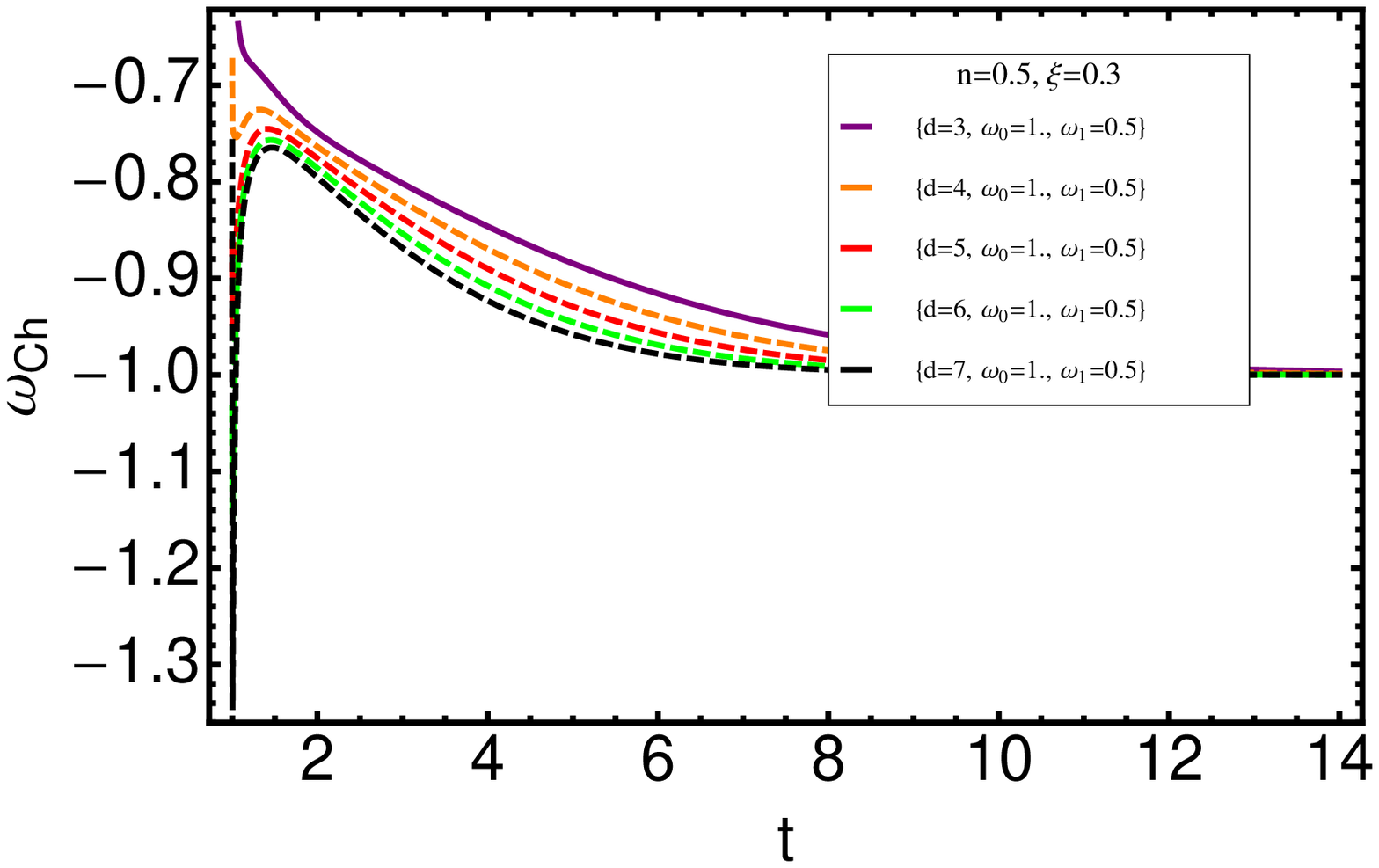} &
\includegraphics[width=50 mm]{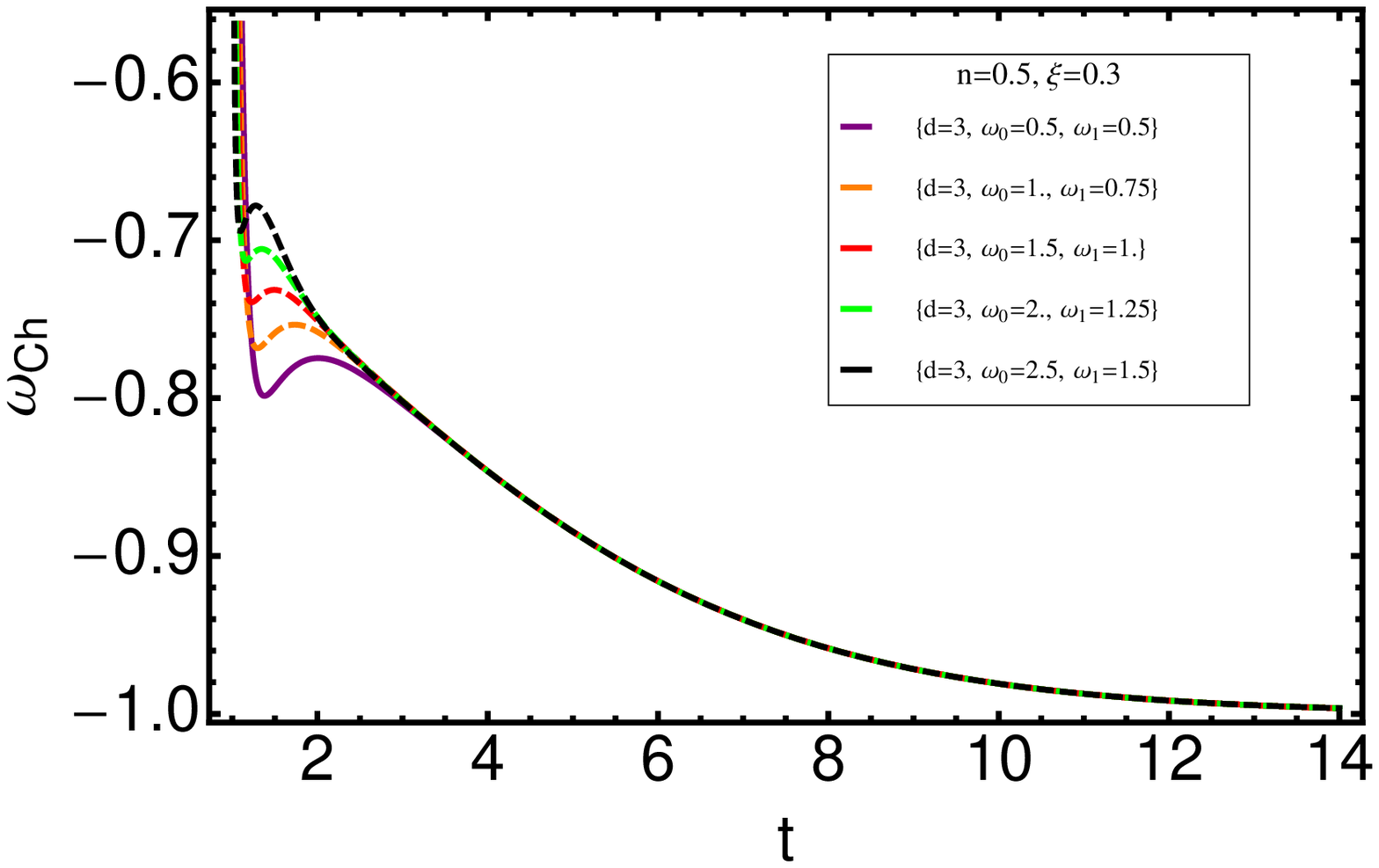}&
\includegraphics[width=50 mm]{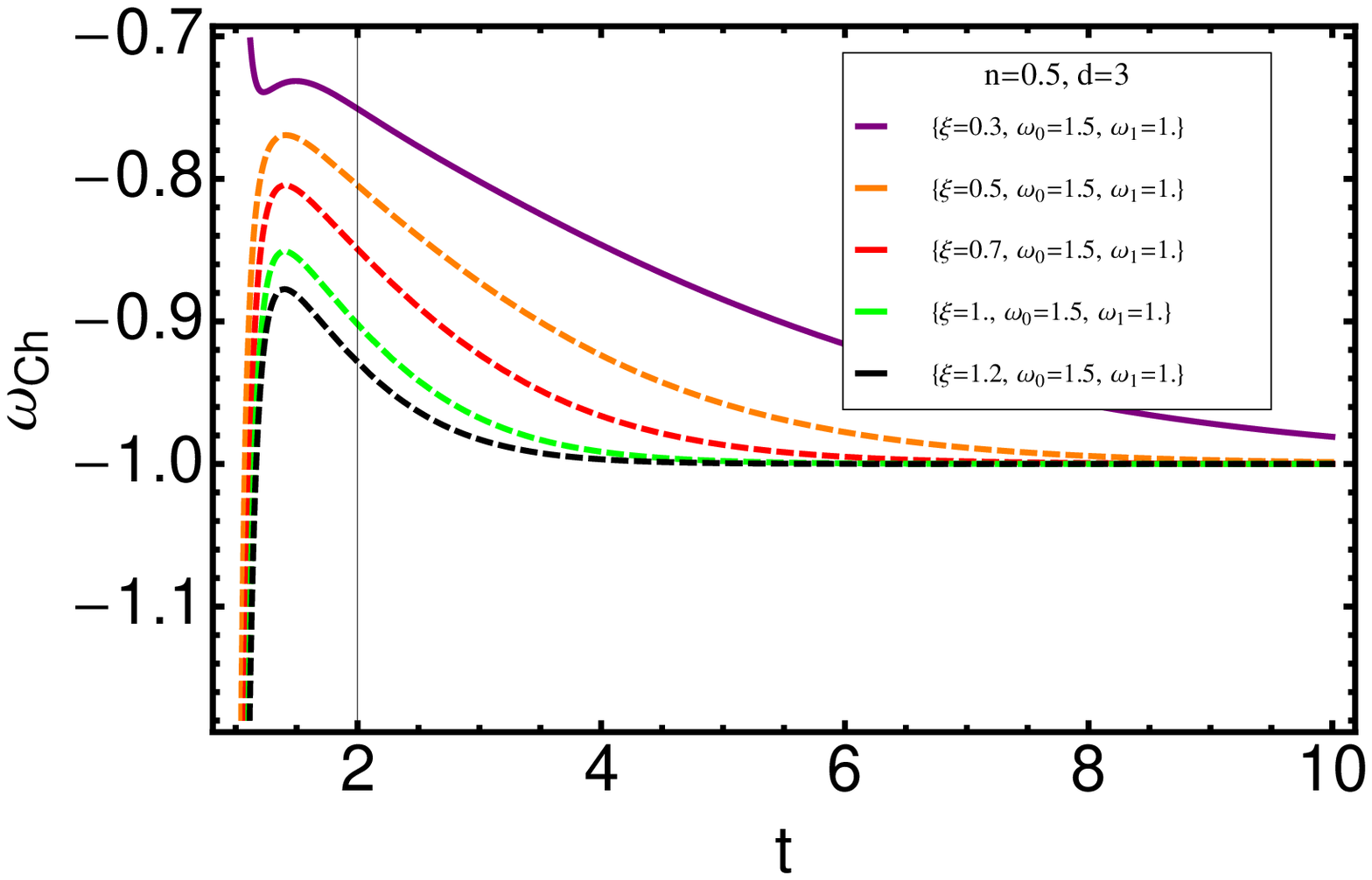}
 \end{array}$
 \end{center}
\caption{Behavior of $\omega_{Ch}$ against $t$. Single fluid model.}
 \label{fig:3}
\end{figure}

\subsection{\large{The second model}}
In this section we consider an effective interacting Ghost DE and Viscous varying modified Chaplygin gas fluid and consider dynamics of the background of the Universe numerically. An interaction may takes the form of the Eq. (1), which is a nonlinear function according to the energy densities of the components. For the comparison of the effect related to the form of interaction $Q$, we present and analyze deceleration parameter $q$ and EoS parameter of the interacting fluids. Also, future investigation showed that considered forms for interaction term $Q$ within this model give equivalent contribution. The dynamics for the energy density components in presence of an interaction $Q$ can be rewritten in the following forms,
\begin{equation}\label{eq:Chgas}
\dot{\rho}+(3+d)\left (1+A-\frac{B(t)}{\rho^{n+1}} \right )H\rho=(3+d) \gamma  \theta \frac{\rho }{\rho+\theta H }H^{2}+(3+d)^{2}\xi H^{2},
\end{equation}
and,
\begin{equation}\label{eq:Gpres}
P_{G}=-\gamma \theta \frac{\rho}{\rho+\theta H}H-\frac{\theta \dot{H}}{(3+d)H}-\theta H.
\end{equation}
On the other hand the dynamics of energy density for the fluid and pressure for ghost dark energy with the interaction term (2) reads as,
\begin{equation}\label{eq:Chgas2}
\dot{\rho}+(3+d)\left (1+A- \gamma-\frac{B(t)}{\rho^{n+1}} \right ) H\rho= \theta \gamma  (3+d) H^{2}+ (3+d)^{2}\xi H^{2},
\end{equation}
and,
\begin{equation}\label{eq:Gpres2}
P_{G}=-\gamma \rho -\frac{\theta \dot{H}}{(3+d)H} +(\gamma -1)\theta H.
\end{equation}

\begin{figure}[h!]
 \begin{center}$
 \begin{array}{cccc}
\includegraphics[width=50 mm]{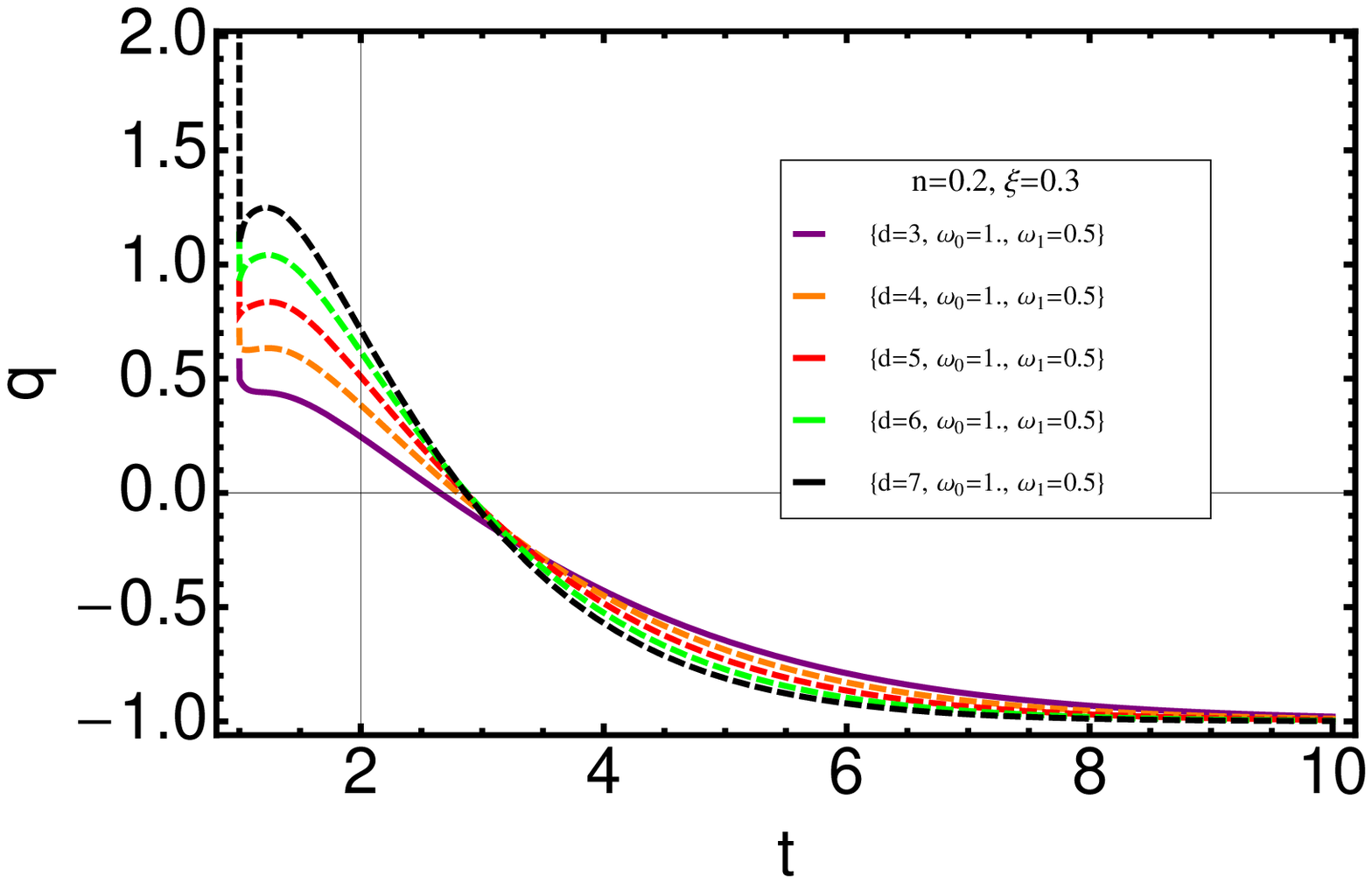} &
\includegraphics[width=50 mm]{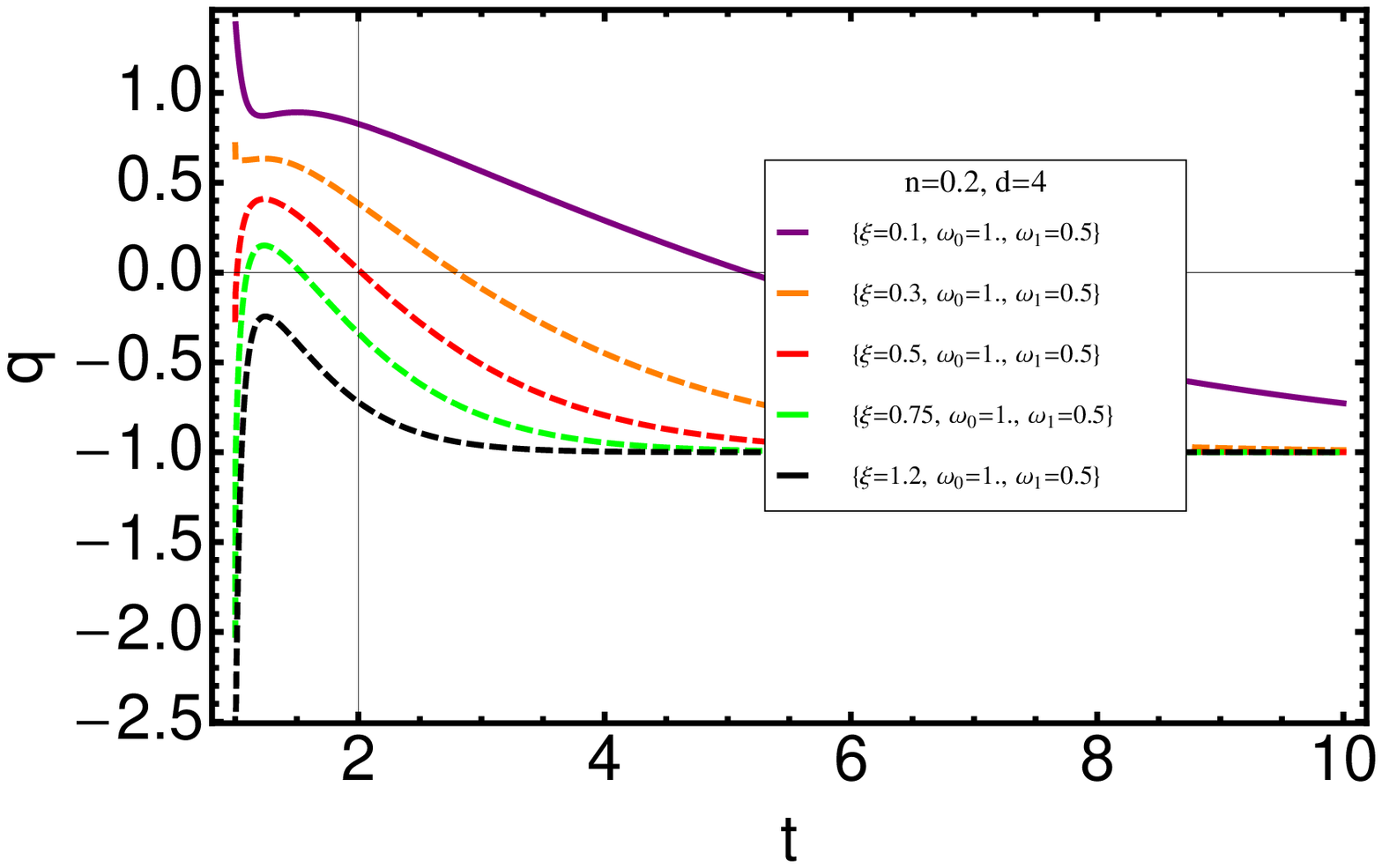}
 \end{array}$
 \end{center}
\caption{Behavior of $q$ against $t$. The second model with $\gamma=0.01$.}
 \label{fig:4}
\end{figure}

In Fig. 4 we draw deceleration parameter in terms of time. We can see that appropriate choice of $\xi<0.4$ gives acceleration to deceleration phase transition. The value of $q$ yields to -1 at the late time which is agree with observational data. Two different forms of interaction term given by (1) and (2) yield to the approximately same results.\\
In the Fig. 5 we can see time evolution of total EoS which yields to -1 as expected. As previous, both interaction terms given by (1) and (2) yield to the similar results. In the plots of Fig. 6 we represent ghost dark energy EoS with variation of parameters.

\begin{figure}[h!]
 \begin{center}$
 \begin{array}{cccc}
\includegraphics[width=50 mm]{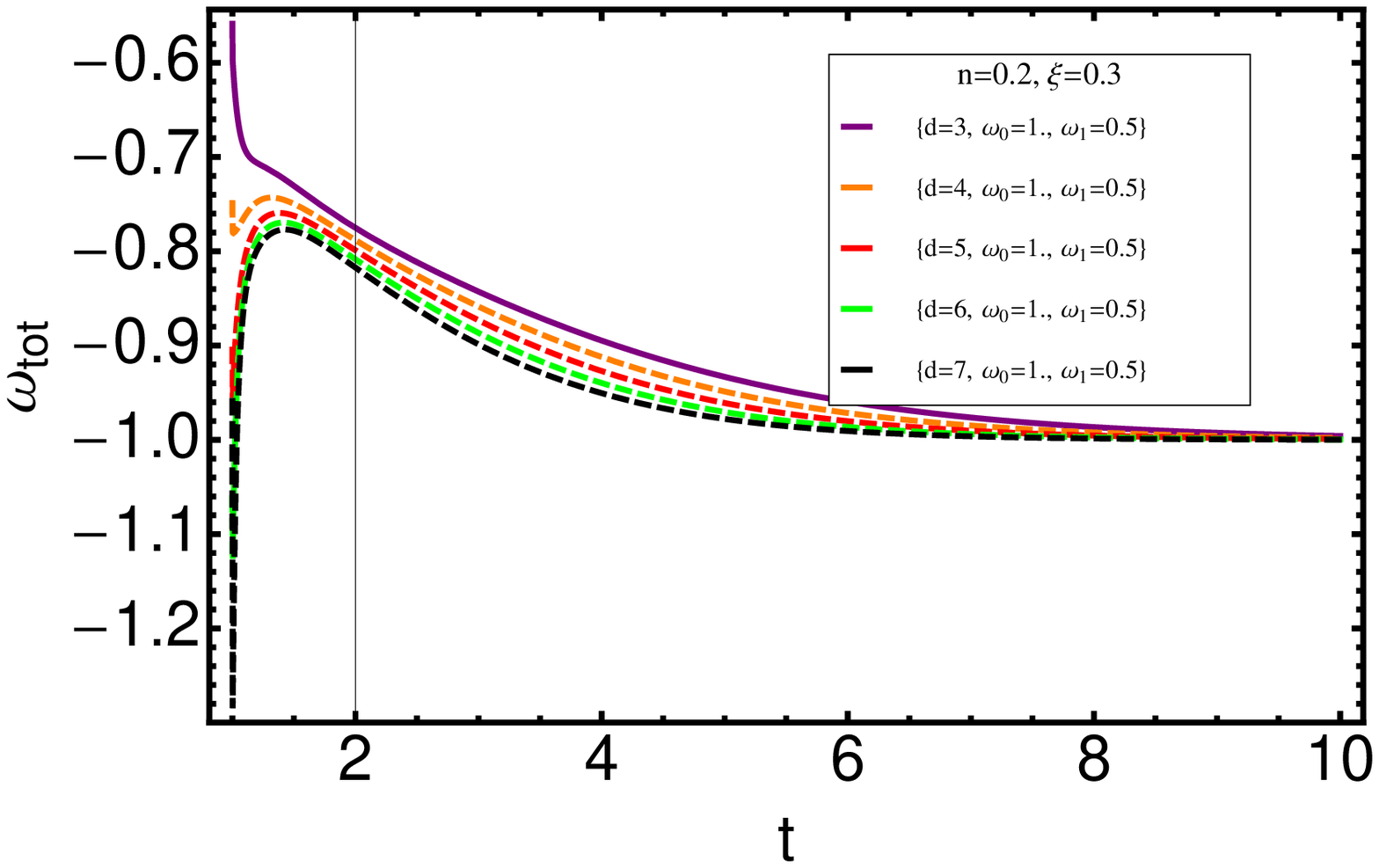} &
\includegraphics[width=50 mm]{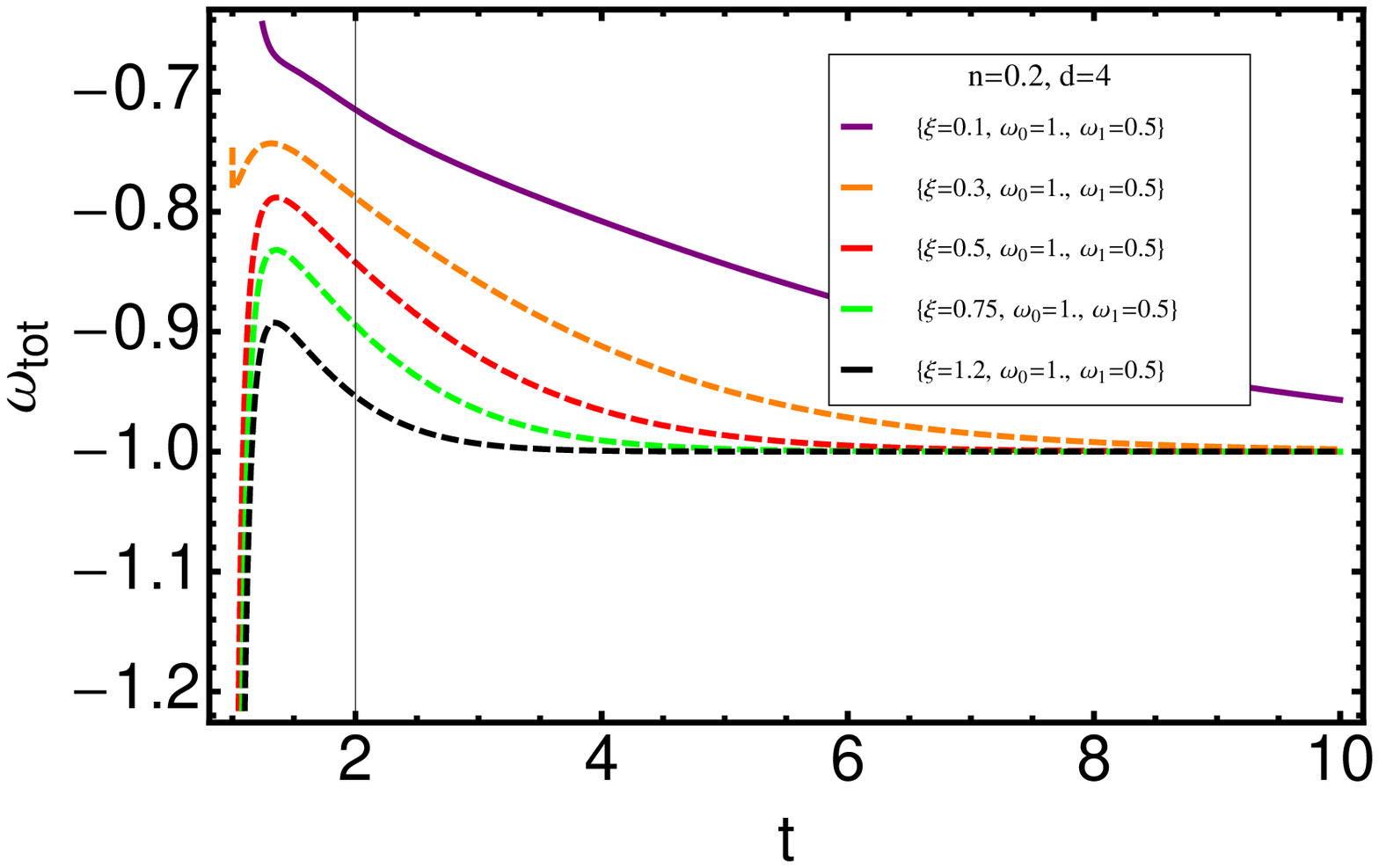}
 \end{array}$
 \end{center}
\caption{Behavior of $\omega_{tot}$ against $t$.  The second model with $\gamma=0.01$. }
 \label{fig:5}
\end{figure}

\begin{figure}[h!]
 \begin{center}$
 \begin{array}{cccc}
\includegraphics[width=50 mm]{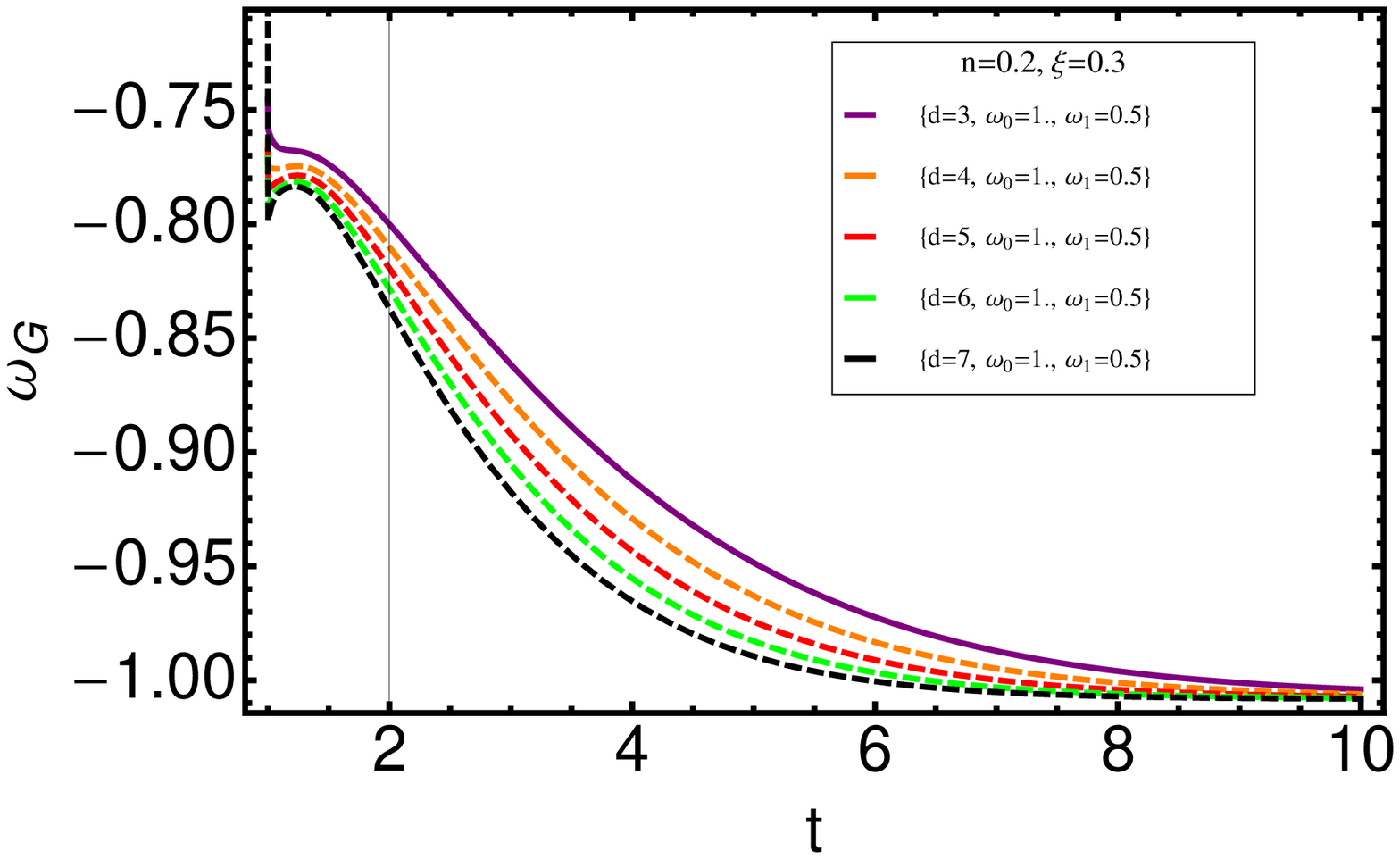} &
\includegraphics[width=50 mm]{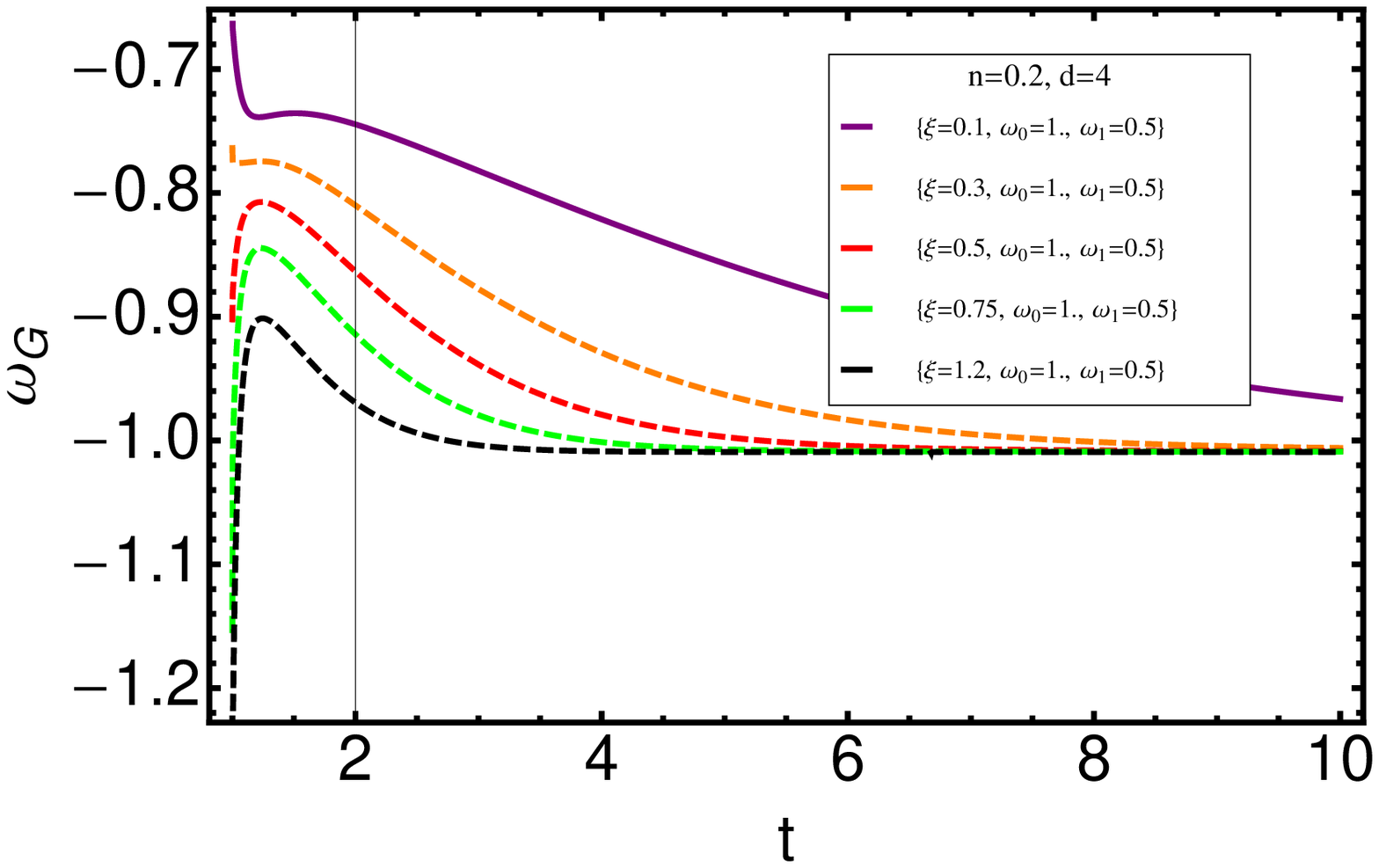}&
\includegraphics[width=50 mm]{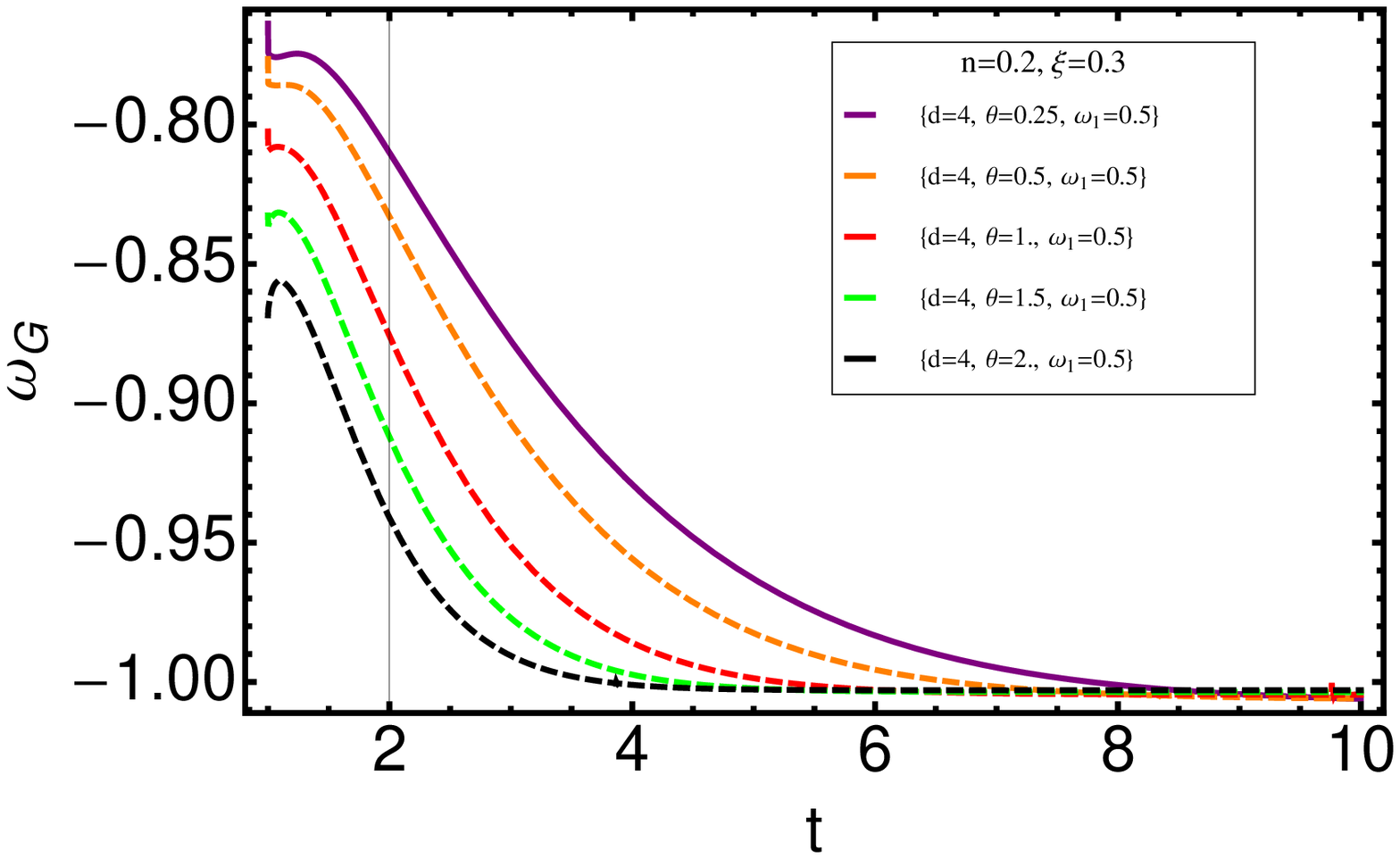}\\
\includegraphics[width=50 mm]{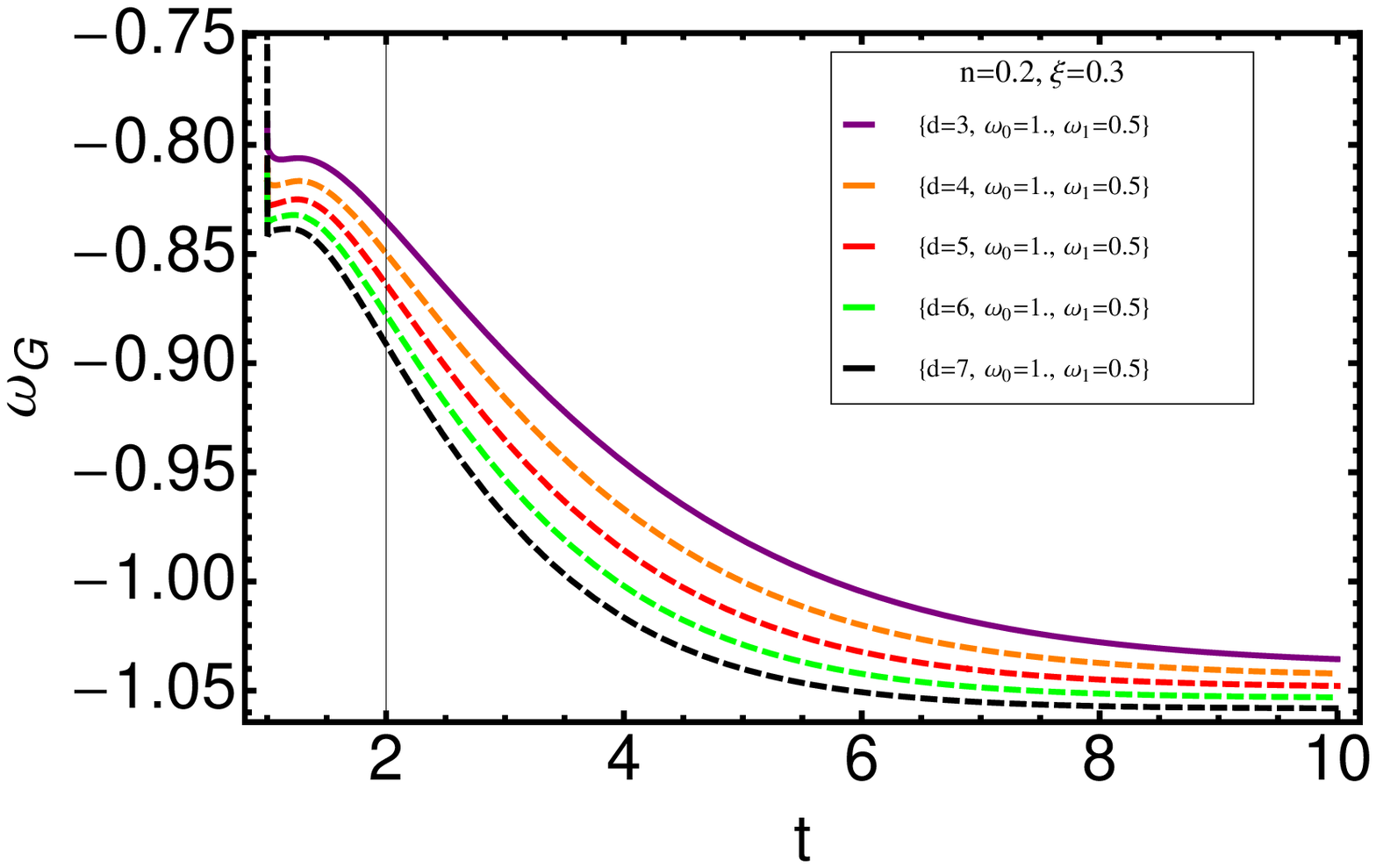} &
\includegraphics[width=50 mm]{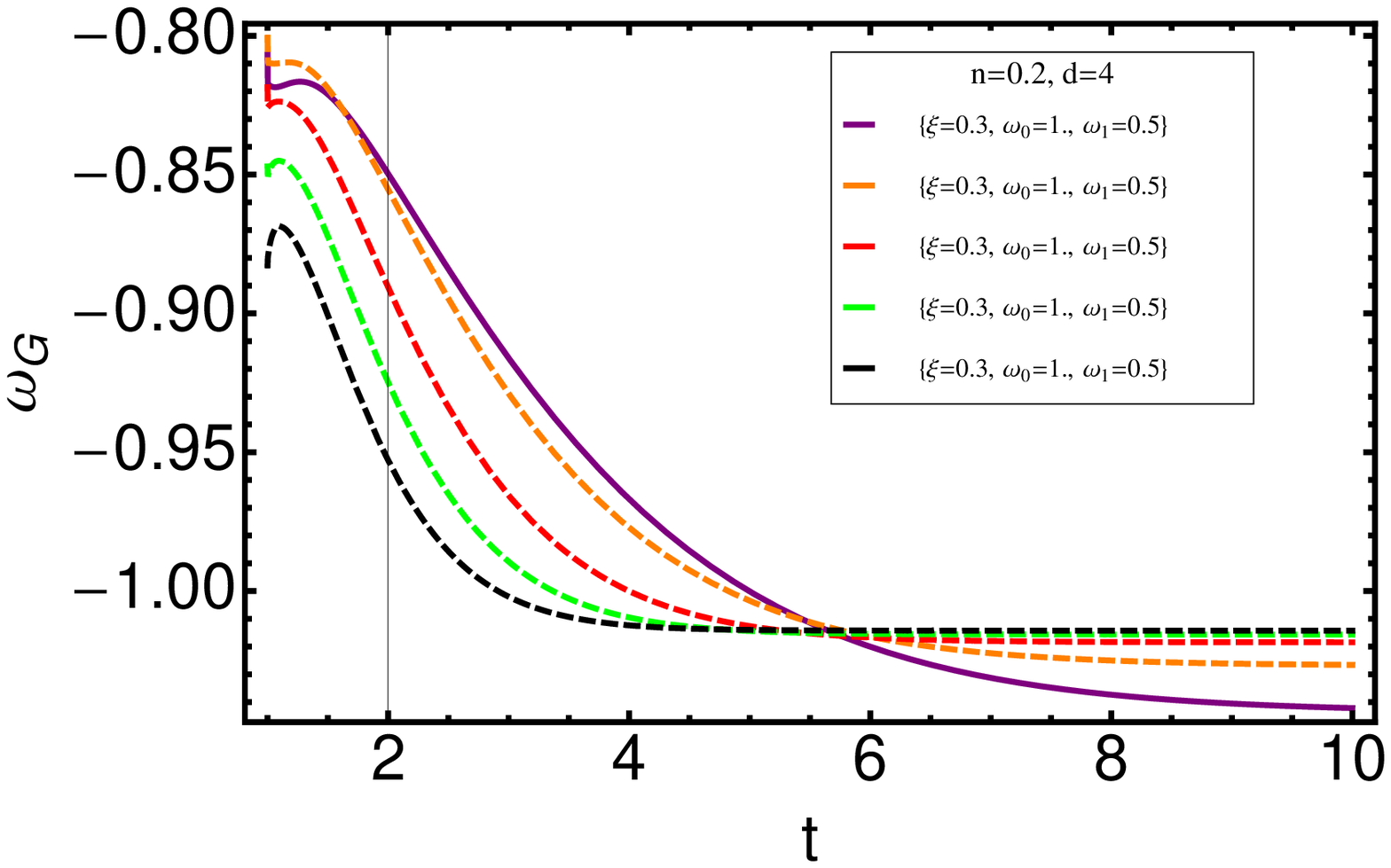}&
\includegraphics[width=50 mm]{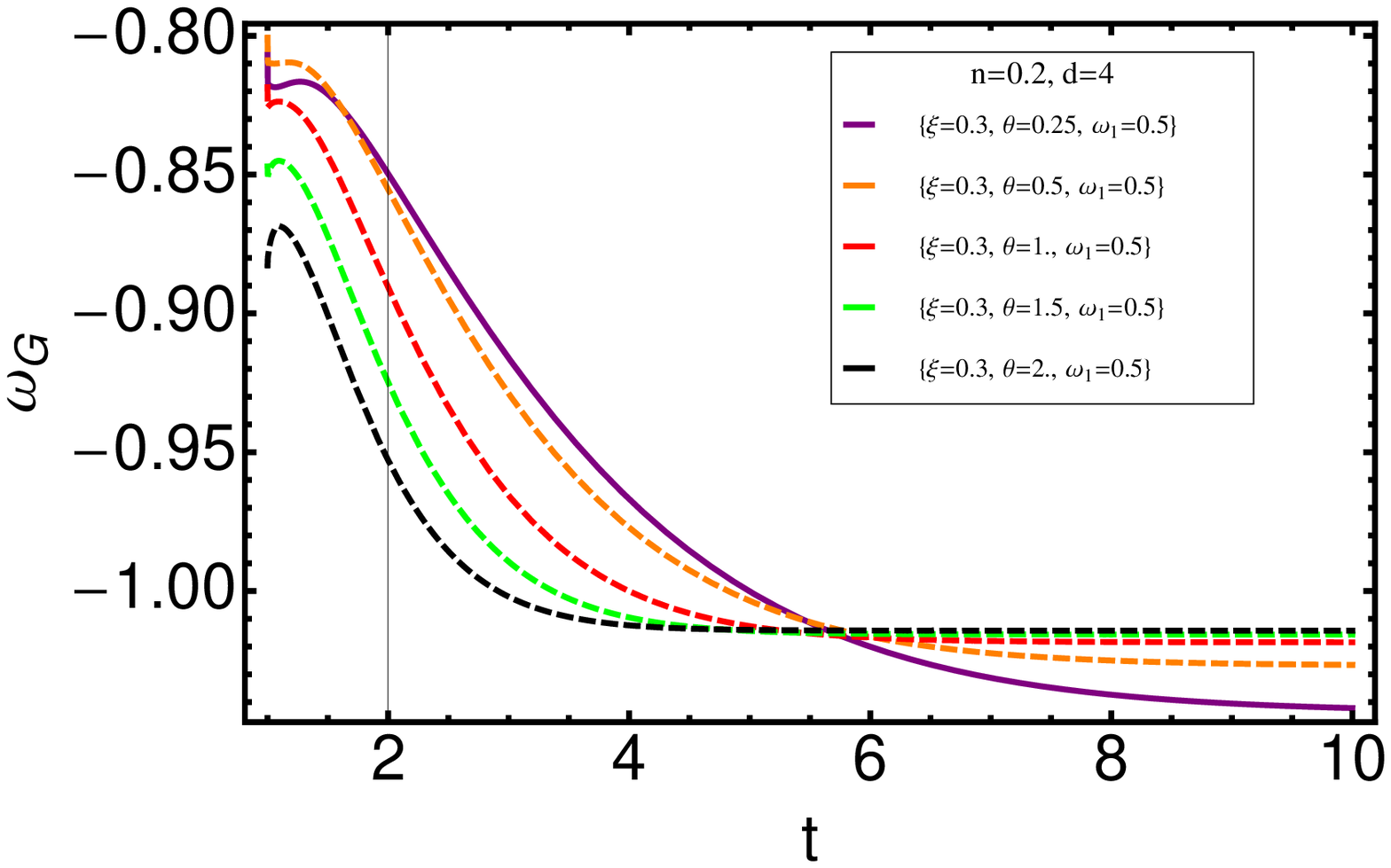}
 \end{array}$
 \end{center}
\caption{Behavior of $\omega_{G}$ against $t$ for $\gamma=0.01$. Top panels correspond to interaction (1). Bottom panels correspond to interaction (2).}
 \label{fig:6}
\end{figure}

\subsection{\large{The third model}}
In this model we consider an interaction between ghost dark energy and viscous modified cosmic Chaplygin gas
and for the dynamics of energy density for Chaplygin gas is described via following differential equation,
\begin{equation}
\dot{\rho}+(3+d)\left (1+A-\frac{1}{\rho^{n+1}}\left [  \frac{B}{1+\omega} -1 + \left ( \rho^{1+n}-\frac{B}{1+\omega}+1\right )^{-\omega} \right ] \right ) H \rho=(3+d) \gamma  \theta \frac{\rho }{\rho+\theta H }H^{2}+(3+d)^{2}\xi H^{2},
\end{equation}
while pressure for ghost dark energy can be obtained from Eq. (\ref{eq:Gpres}).\\
In Fig. 7 the evolution of Hubble expansion parameter specified which show that is decreasing function of time and yields to a constant at the late time.\\
Also, in Fig. 8 we draw deceleration parameter and see appropriate behavior comparing with observational data.

\begin{figure}[h!]
 \begin{center}$
 \begin{array}{cccc}
\includegraphics[width=50 mm]{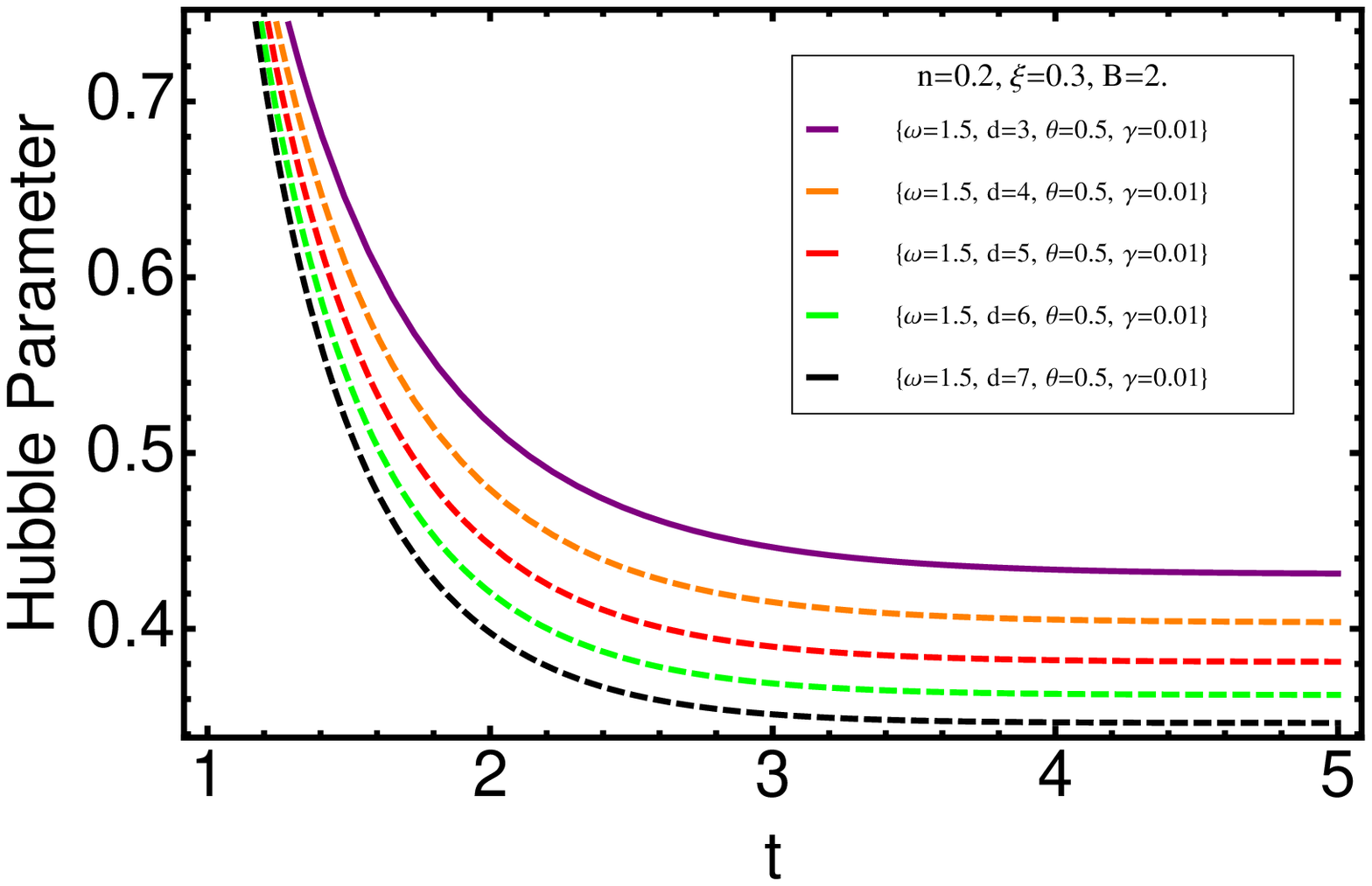}&
\includegraphics[width=50 mm]{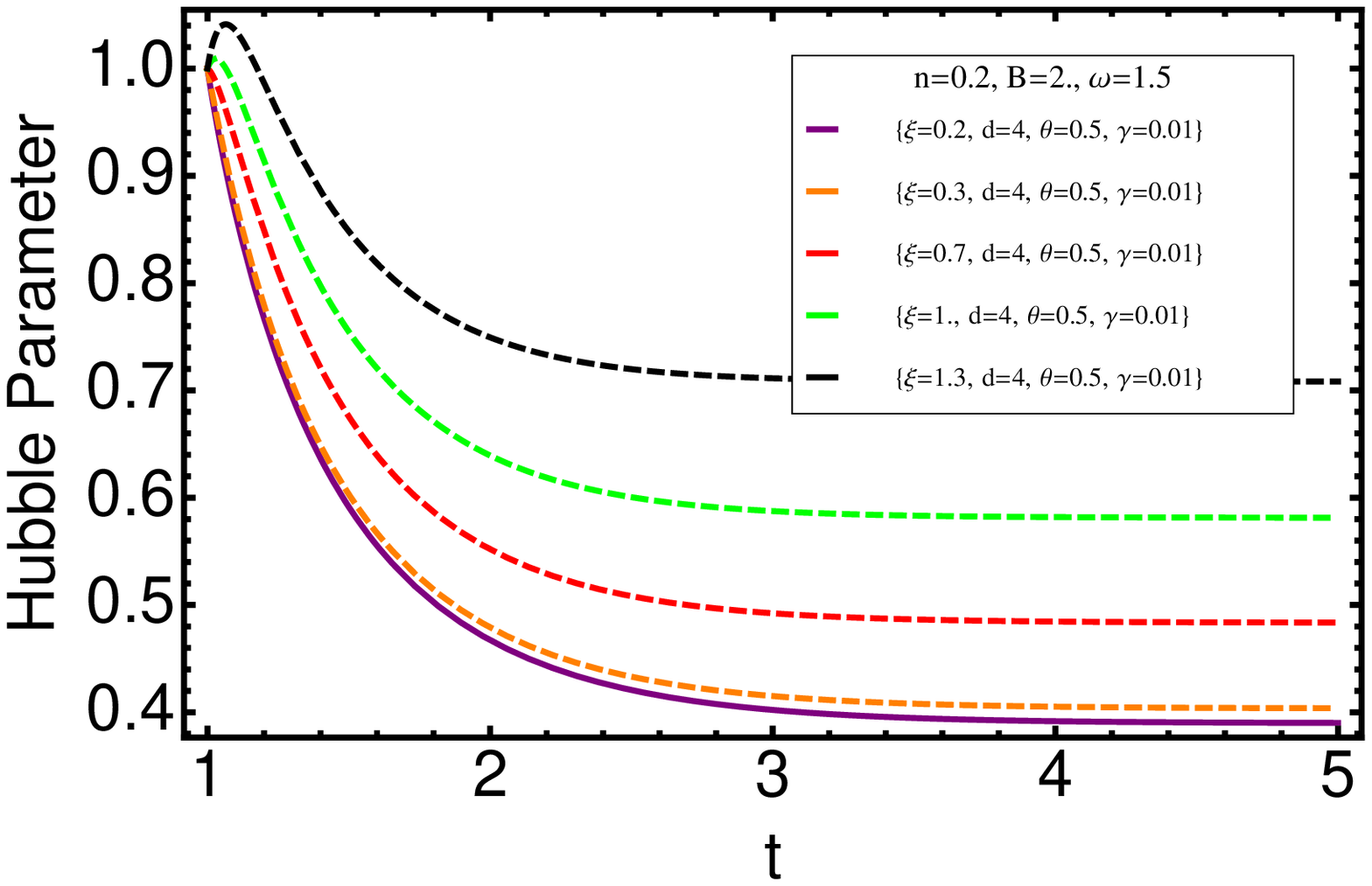}
 \end{array}$
 \end{center}
\caption{Behavior of $H$ against $t$. Model 3.}
 \label{fig:7}
\end{figure}

\begin{figure}[h!]
 \begin{center}$
 \begin{array}{cccc}
\includegraphics[width=50 mm]{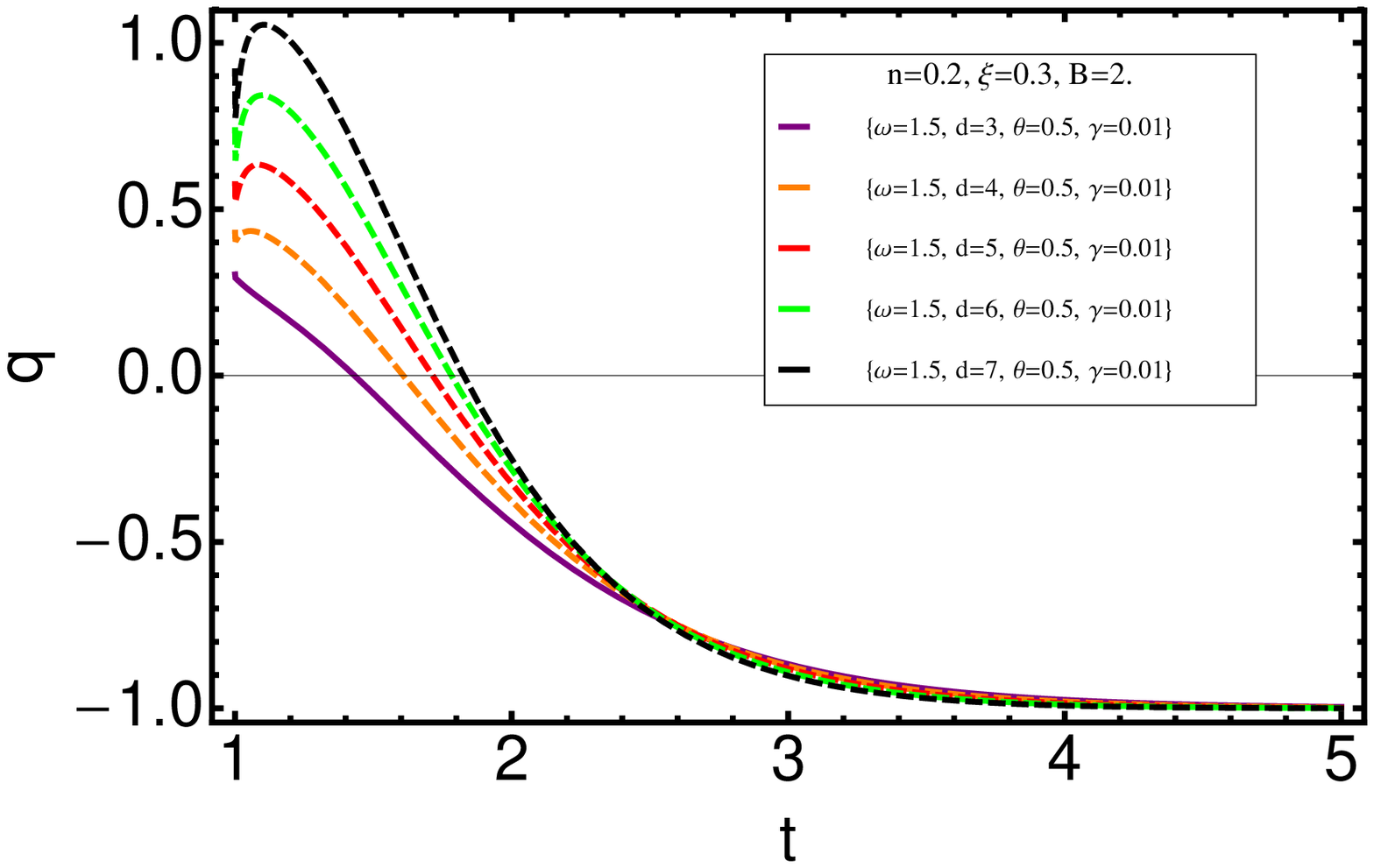} &
\includegraphics[width=50 mm]{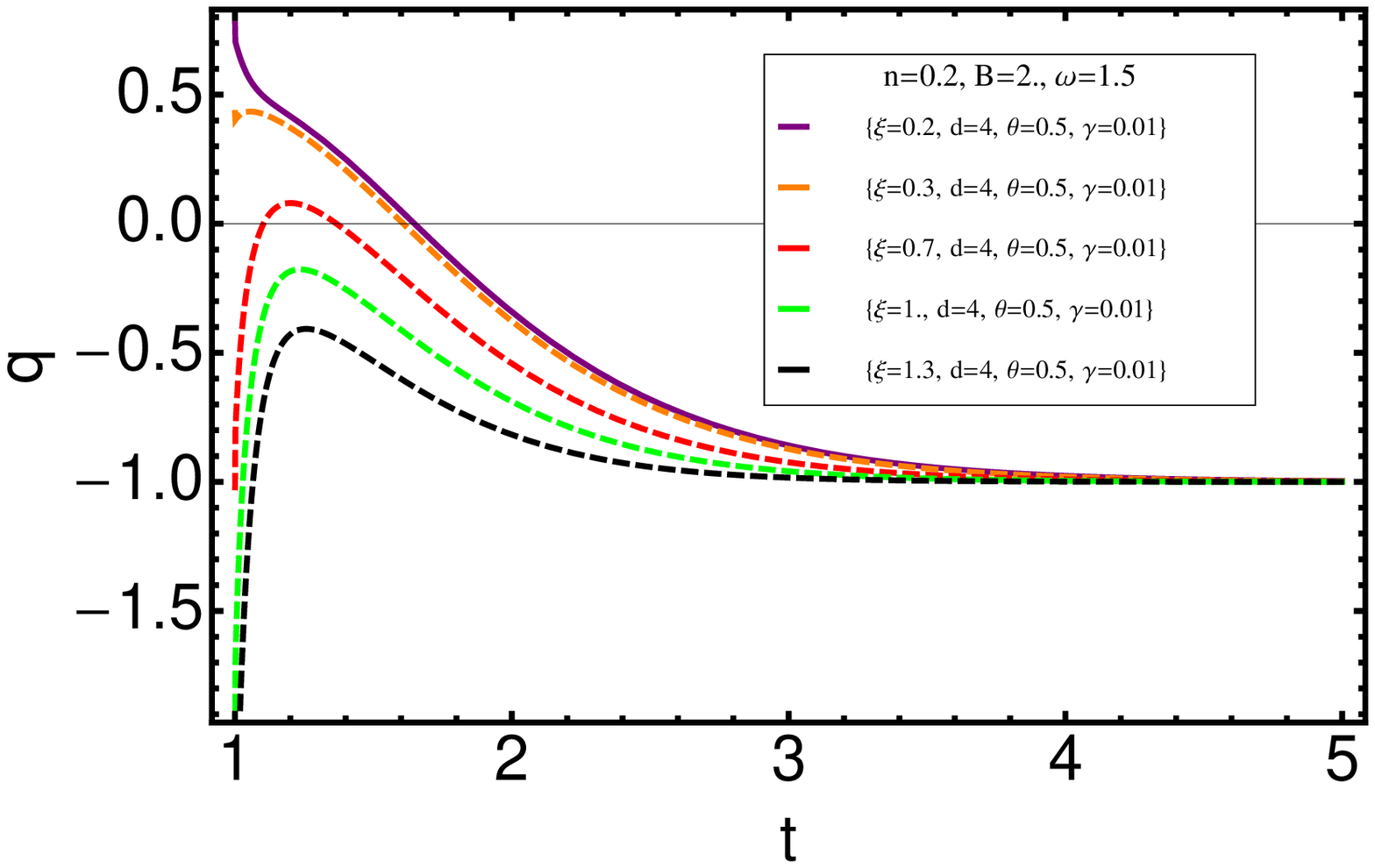}
 \end{array}$
 \end{center}
\caption{Behavior of $q$ against $t$. Model 3.}
 \label{fig:8}
\end{figure}

In Fig. 9 we can see behavior of total EoS by variation of viscous parameter and number of extra dimensions. We can see that the value of $\omega_{tot}$ yields to -1 at the late time. Just there are special cases with $d=3$ and low viscous parameter. In these cases total EoS is totally decreasing function of time but in the cases of $d>3$ the value of $\omega_{tot}$ increases first to a maximum and then decrease to yields -1. Similar behavior happen for $\xi\leq0.2$.

\begin{figure}[h!]
 \begin{center}$
 \begin{array}{cccc}
\includegraphics[width=50 mm]{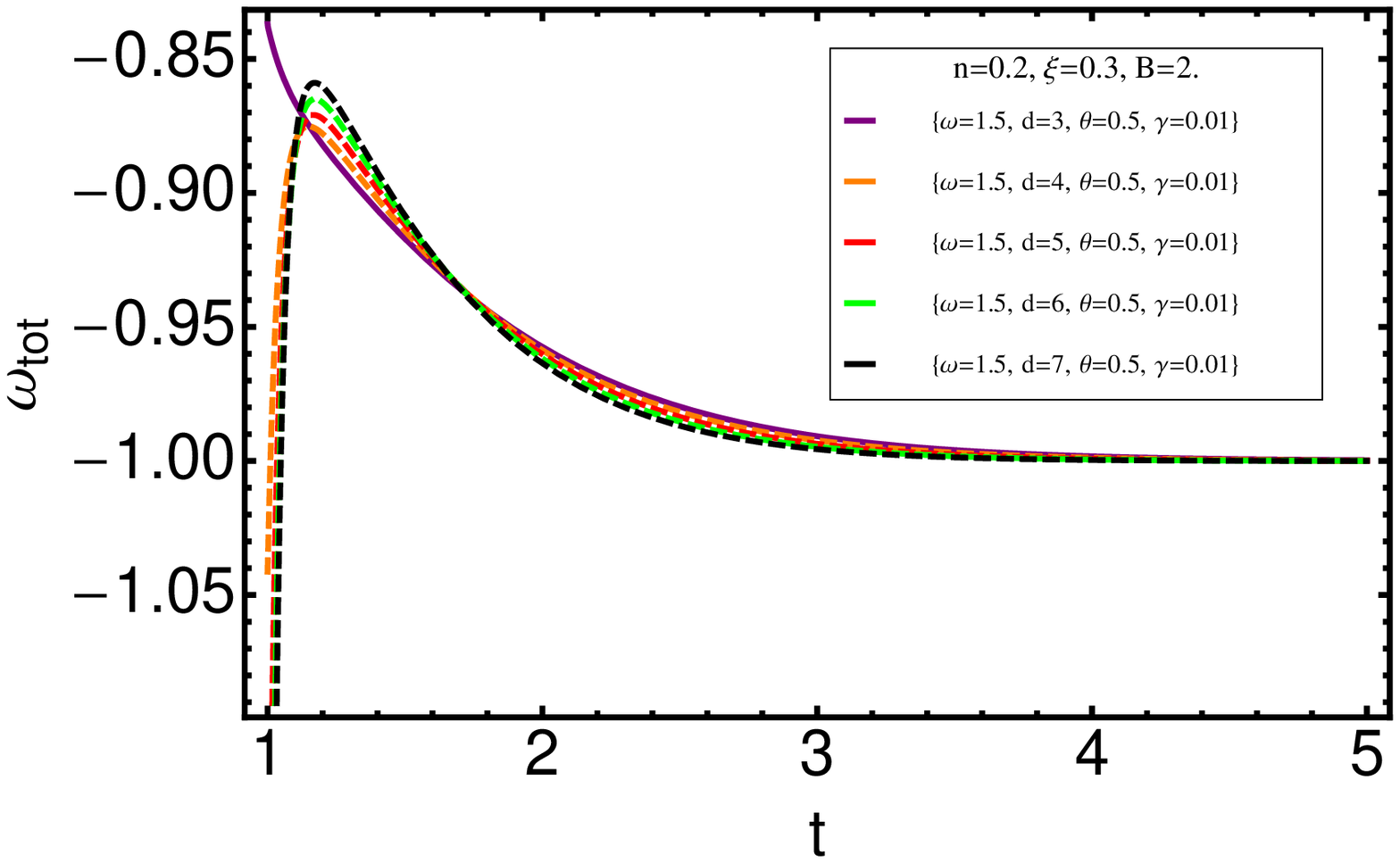} &
\includegraphics[width=50 mm]{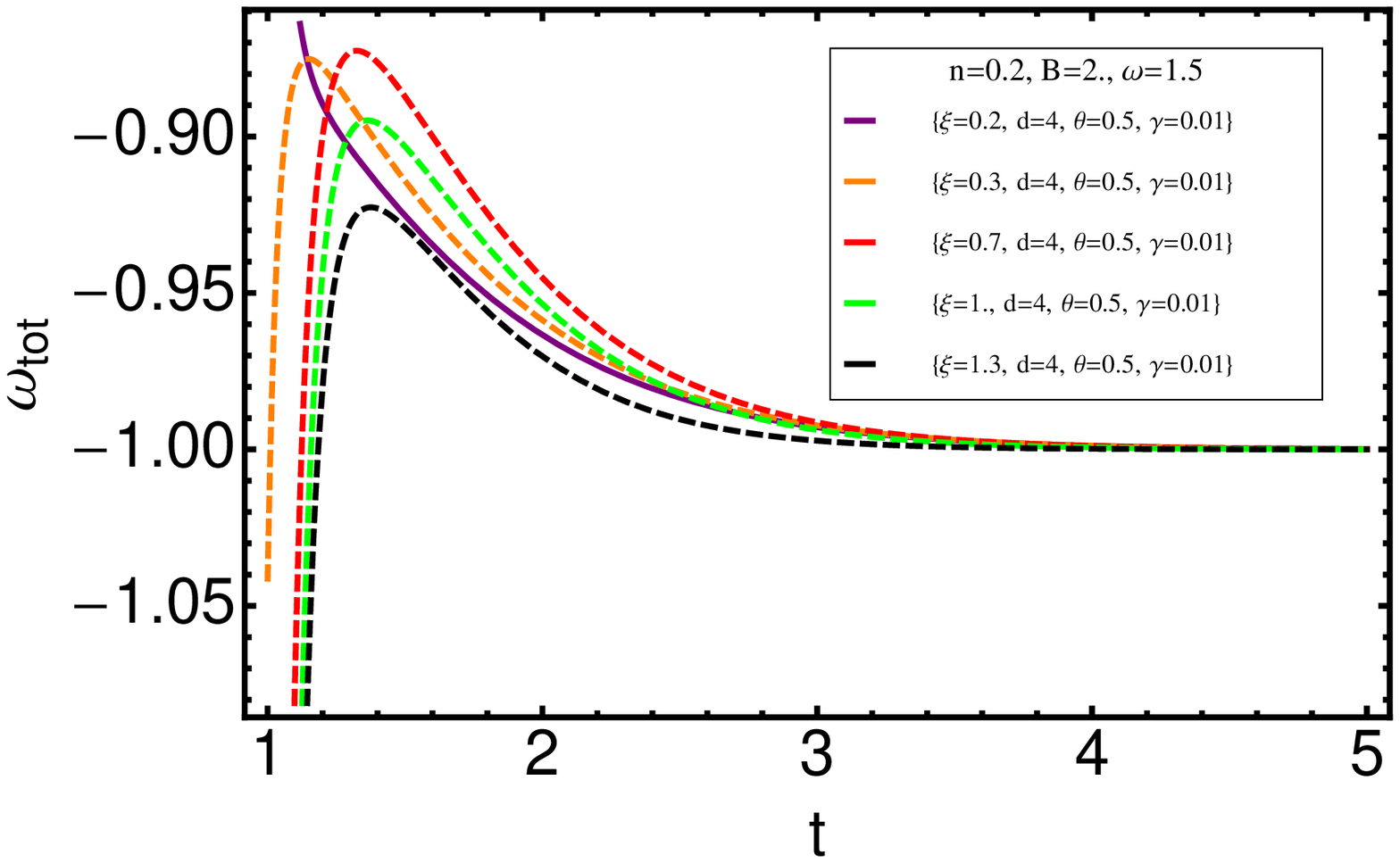}
 \end{array}$
 \end{center}
\caption{Behavior of $\omega_{tot}$ against $t$. Model 3.}
 \label{fig:9}
\end{figure}

\section{\large{Discussions}}
In this paper, we considered two-component fluid with extra dimensions as a toy model for the universe. For the first component we considered two different models of Chaplygin gas which were viscous varying modified Chaplygin gas and viscous modified cosmic Chaplygin gas. The second component assumed as ghost dark energy. We also considered possibility of interaction between components by two special cases of interaction term. First of all we studied an special case of single fluid as viscous varying modified Chaplygin gas. Then, we studied two models of interacting two-component fluid. By using numerical analysis we obtained behavior of some important cosmological parameters such as Hubble expansion, deceleration and EoS parameters. We found that all models have agreement with observational data.

\end{document}